# Low-energy run of Fermilab Electron cooler's beam generation system

Lionel Prost, Alexander Shemyakin
*Fermi National Laboratory, Batavia, IL*
Alexei Fedotov, Jorg Kewisch
*Brookhaven National Laboratory, Upton, NY*

*Abstract*

As a part of a feasibility study of using the Fermilab Electron Cooler for a low-energy Relativistic Heavy Ion Collider (RHIC) run at Brookhaven National Laboratory (BNL), the cooler operation at 1.6 MeV electron beam energy was tested in a short beam line configuration. The main result of the study is that the cooler beam generation system is suitable for BNL needs. In a striking difference with running 4.3 MeV beam, no unprovoked beam recirculation interruptions were observed.

## I. Introduction

Electron cooling was proposed to increase luminosity of the RHIC collider for heavy ion beam energies below 10 GeV/nucleon. Providing collisions at such energies, termed RHIC "low-energy" operation, will help to answer one of the key questions in the field of QCD about the existence and location of a critical point on the QCD phase diagram. The electron cooling system should deliver an electron beam of required good quality over the energies of 0.9-5 MeV [1.1].

Apart from some modifications needed to address electron beam transport at low energies and recombination suppression in the cooling section, it appears that the existing Recycler's electron cooler at FNAL is ideally suited for low-energy RHIC cooling. An evaluation of modifications needed to use the Recycler's Pelletron for RHIC is under study. As part of these studies, experimental time was requested to explore Pelletron operation at low beam energies relevant for RHIC. The goal of these experiments was to understand at what values of magnetic field on the cathode and what electron beam currents it is feasible to operate the Pelletron at such energies. Because of operational restrictions, the study was performed with a short beam line (so called U-bend). Operation in this line allows one to analyze issues related to the generation of the electron beam in the energy recovery mode and to gain experience with the beam transport at lower beam energy.

As of the summer of 2010, the lowest energy at which RHIC was able to provide colliding beams for physics was at $\gamma$ = 4.1. Possibility of RHIC operation at even lower energies is the subject of future studies. Therefore, for the present studies of the Pelletron, the electron beam kinetic energy of 1.6MV was chosen, which corresponds to $\gamma$ = 4.1. The first step of the study was to start with the nominal magnetic field of 90G on the cathode and establish recirculation for the highest DC current possible. The next step was to increase the magnetic field on the cathode to 250G and again establish recirculation of high electron beam current. Although the original design of the electron cooler for RHIC required a rather modest electron beam current of about 60-100mA, recent modifications to the design would benefit from increasing the electron beam current up to 200mA. This report summarizes simulations which were carried out in preparation for these studies and our findings and observations in the course of the measurements.

# II. Setup

The Pelletron [2.1] is a 6MV electrostatic accelerator which works in the energy recovery mode. It consists of two ‘acceleration’ columns (actually, one column accelerates the beam and the second decelerates it) composed of 12 accelerating tubes each, contained in a pressure vessel filled with pressurized $SF_6$ gas (~70 psi). There are 6 ‘levels’ to ground, each of which contains two focusing solenoids (one for each column) each with a pair of dipole correctors. The electron gun is embedded in a magnetic field and can generate a few Amperes electron beam with low thermal emittance. The collector is water-cooled and recuperates the beam which has been decelerated to 3 kV. The deposited power in the collector is of the order of a few kilowatts (depending on the beam current).

A drawing of the Pelletron in the short beam line configuration (U-Bend mode) is shown on Figure 2.1. In this configuration, the 90° bends (each bend is composed of two 45° bend dipoles, in pink on Figure 2.1) are turned off such that the beam generated in the gun (top left) goes straight down to the 180° bend (U-Bend) and returns to the collector (top right). Not shown on the drawing are two ‘bucking coils’ which compensate the remnant fields from the bends, which are strong enough to deflect the beam in a way that the correctors cannot correct.

One important feature for tuning purposes is the lack of Beam Position Monitors (BPMs) inside the acceleration and deceleration columns. The first BPM (A05) is located right at the exit of the acceleration column; the last BPM (D05) just before the entrance of the deceleration tube. There are two diagnostics stations with Optical Transition Radiation (OTR) monitors: one located just below the first 45° bend dipole; the second on the other side of the U-Bend, between the two long solenoids (Walker solenoids). The OTR monitors were not used during this study but were useful during the initial commissioning of the Pelletron. A ‘crash scraper’ is located a little bit downstream of the U-Bend. In conjunction with the protection system, it helps prevent full discharges.

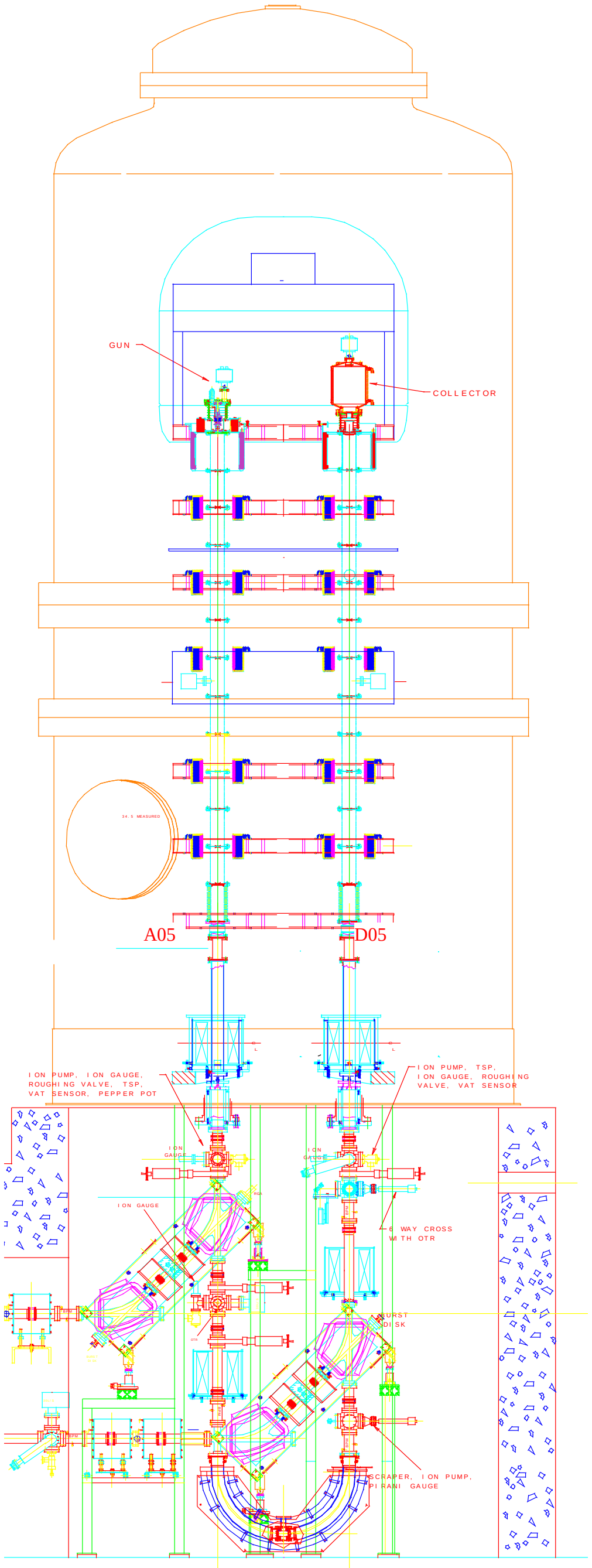

**Figure 2.1.** CAD drawing of the Pelletron in the U-Bend configuration.

# III. Simulations

In order to prepare running the Pelletron at low energy (with respect to its nominal operation energy of 4.3 MV), simulations of the beam dynamics in the Pelletron and short transport lines (U-bend configuration) were carried out with the goal of obtaining initial settings for the focusing elements. Two codes were employed: the SAM/UltraSAM/BEAM package (often called SAM below for simplicity) [3.1, 3.2, 3.3] for gun calculations and beam transport simulations in the acceleration column at the beginning of acceleration; the OptiM code [3.4, 3.5] for optics calculations for the rest of the beam line.

The SAM/UltraSAM code allows simulating intense particle beam dynamics with space-charge effects. It also allows simulating particle emission from surfaces (using a space-charge limited model). The BEAM code evaluates the dynamics of cylindrical steady-state beams of charged particles in long beam lines, taking into account the beam transverse self fields. SAM and BEAM are complementary in the sense that SAM/UltraSAM results are easy to export into BEAM, which uses them as input parameters for further calculations.

OptiM is a user-friendly, matrix-based optics code which is used when the more detailed and computing intensive SAM package is no longer advantageous and/or when the axial-symmetry of the beam line is broken (*e.g.*: dipole bending magnets), which SAM cannot accommodate (at least not concurrently with an otherwise axially-symmetric beam line).

Three sets of calculations were performed [3.6]:

$U_{pell}$ = 1.6 MV with $U_a$ = 10 kV, $I_b$ = 100 mA and ~90 G on the cathode (Set A);
$U_{pell}$ = 1.6 MV with $U_a$ = 10 kV, $I_b$ = 100 mA and ~250 G on the cathode (Set B);
$U_{pell}$ = 1.6 MV with $U_a$ = 20 kV, $I_b$ = 700 mA and ~250 G on the cathode (set C).

$U_{pell}$, $U_a$ and $I_b$ are the Pelletron terminal voltage (and beam energy), the anode voltage and the beam current, respectively. The primary choice of 10 kV for the anode voltage was motivated by the focusing effect that the beam experiences when it exits the gun. At a lower anode voltage, the excursion of the beam envelope at the beginning of the accelerator column, on which the magnetic lens (so-called gun solenoid) has a limited effect, is smaller than for higher anode voltages, for a given beam current. On the other hand, the lower anode voltage limits the maximum current that can be extracted. Since originally, the scope of the study was to recirculate a 100 mA beam, $U_a$ = 10 kV was adequate and preferable to limit the beam size in the low energy part of the acceleration column. Later on, the scope of the study was extended to trying to recirculate a higher beam current, which required an increase of the anode voltage. The choices for the magnetic field at the cathode reflect both hardware limitations and foreseen beam requirements in the cooling section for RHIC.

With SAM, the first goal was to find lens settings for which the calculated beam envelope in the acceleration column would easily clear the electrodes and limit large envelope variations (*i.e.* limit large max to min ratios). Then, for Sets A&B, the settings were checked to be adequate in the case of a full discharge *i.e.* the beam envelope would still be sufficiently far away from the electrodes in case of a sudden anode voltage drop.

With OptiM, the rest of the beam line was modeled and settings adjusted so that the envelope would remain far from the vacuum chamber. In addition, attempts were made to lower the dispersion in the deceleration column as much as possible. According to the previous experience, configuration with maximum dispersion right after the bend, where the crash scraper is located, is efficient for preventing full discharges in cases of Pelletron voltage drops caused by a beam loss or corona currents in the tank.

The following sets of plots correspond to the final settings we arrived at after multiple iterations. They were used as a starting point for tuning during the studies. Figure 3.1 shows three current density profiles obtained at the cathode.

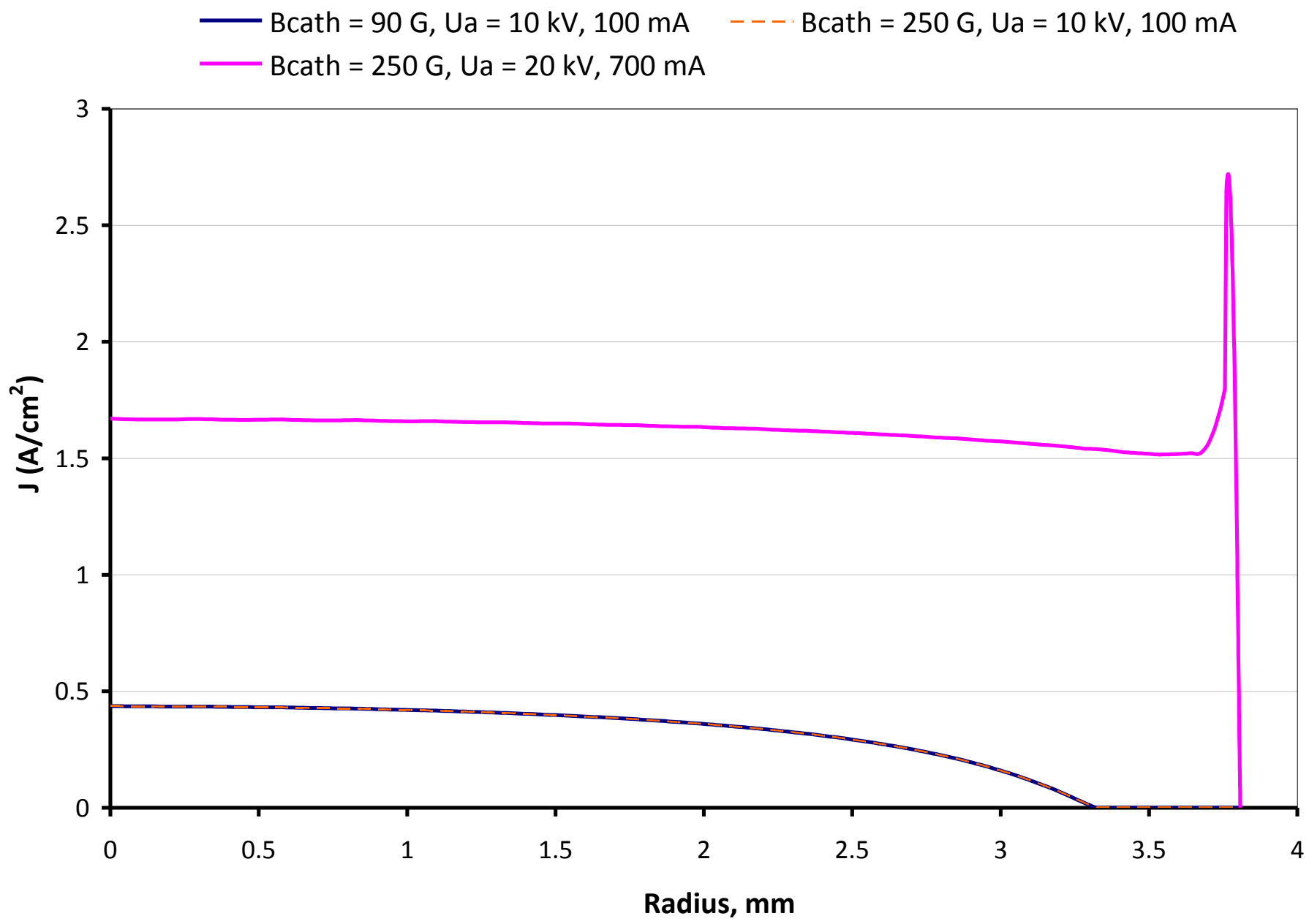

**Figure 3.1.** Current density distribution at the cathode for all 3 cases. Since the only difference between Set A and B is the magnetic field on the cathode, the current density distributions are identical.

Because the emitting size of the cathode only depends on the electrostatic field configuration in the gun (and not the magnetic field), the current density distributions at the cathode are the same for Sets A & B. However, the effective emittance, dominated by the angular momentum of the beam, is larger in Set B than in Set A. Thus, the gun produces beams with very different characteristics in the two cases. This is illustrated in Figure 3.2 (screenshots of UltraSAM post-processor), which shows the beam evolution in the gun for all 3 cases.

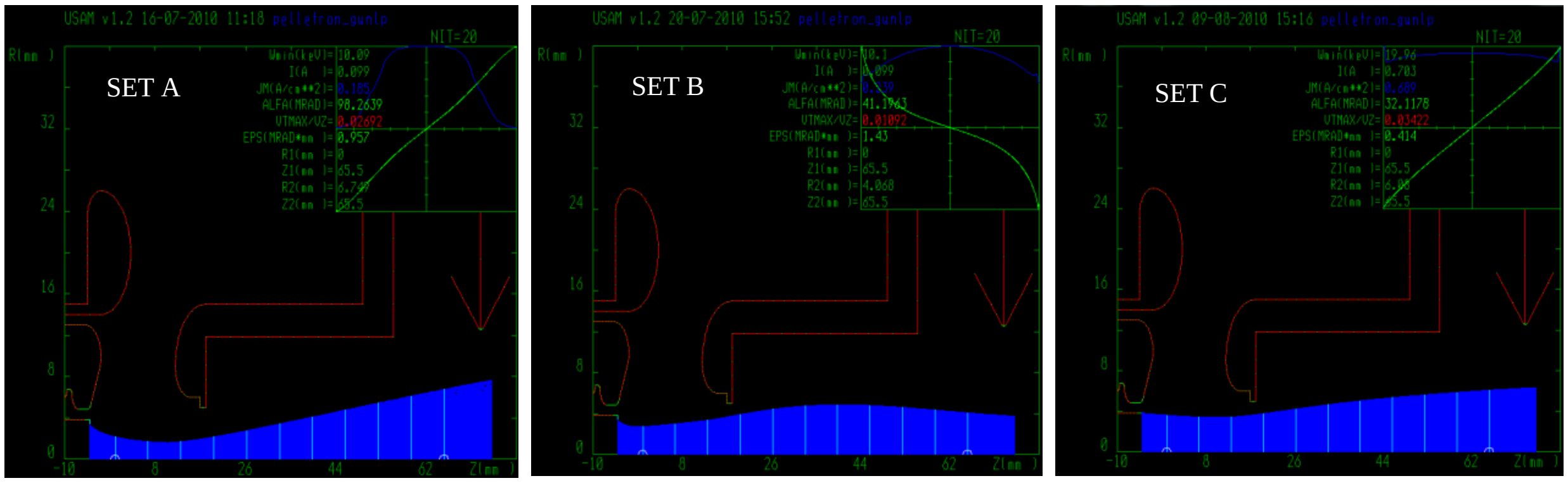

**Figure 3.2.** Gun simulations with SAM/UltraSAM for Sets A (left), B (center) and C (right). In the upper right corner of each plot, the beam phase space that is utilized as input parameters in BEAM simulations is displayed.

Figure 3.3 shows the beam envelope as calculated in BEAM for all three cases on a single plot. The limiting aperture is also indicated by the red line. It represents the 1" diameter diaphragm at the entrance and exit of each individual acceleration tube.

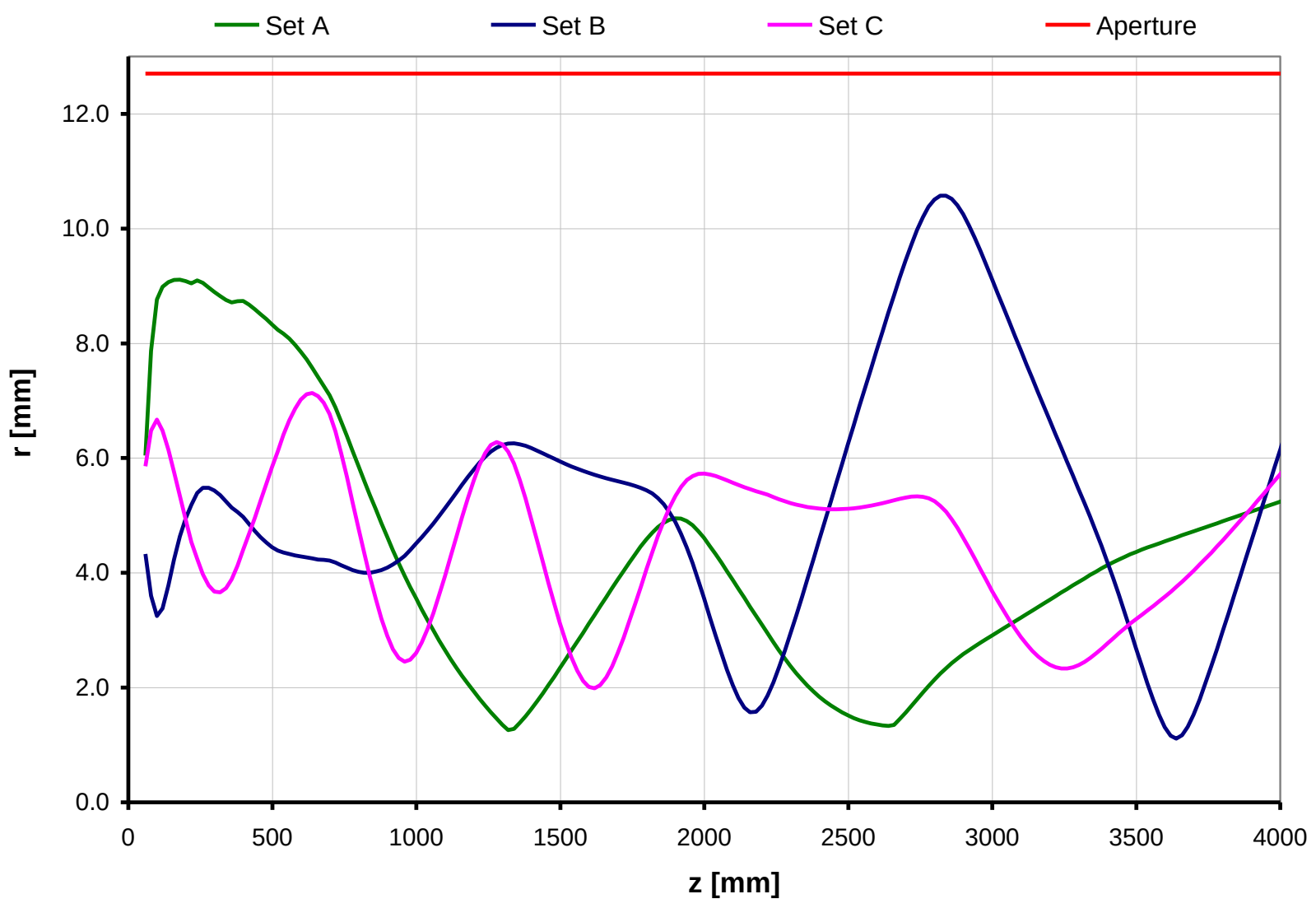

**Figure 3.3.** Beam envelopes for all three cases with intermediate focusing settings. The red line at the top of the plot indicates the aperture limit, which is determined by the position of the inner most electrodes (at the entrance and exit of each individual acceleration tube).

Note that the beam envelopes depictured in Figure 3.3 do not exactly correspond to the settings ultimately chosen and summarized in Table 3.1. BEAM simulations do determine the settings for the first two lenses (SPA00 and SPA01), but, as explained below, although OptiM was used to model the rest of the beam line, it also duplicates the high energy end of the acceleration tube. This is required because in order to accommodate the beam in the transport lines and deceleration tube, changes to the focusing settings must be done as early as the 4$^{th}$ or 5$^{th}$ focusing lens within the acceleration tube. Recalculating the envelope with BEAM each time would be very time consuming and, although the agreement between the envelopes obtained with BEAM and the ones obtained with OptiM is not always very good, the level of accuracy is adequate for tuning purposes.

Figures 3.4 to 3.9 (OptiM screenshots) complete the sets of simulations carried out and show the beam sizes in the beam line from calculations with OptiM as well as the dispersion functions with the final settings from Table 3.1. Note that the initial parameters for OptiM are obtained from the phase picture output from BEAM at z ~ 1000 m (corresponding to a beam energy of ~415 kV). For these calculations, the initial parameters (R and dR/dz) extracted from the phase picture correspond to the very edge of the beam (what could be considered the halo of the beam) and, in most cases, are not a good representation of the core of the beam (because the angle becomes non-linear at the edge). Nevertheless, in this case, since we are interested in avoiding any beam loss, the behavior of the 'halo' is what matters most.

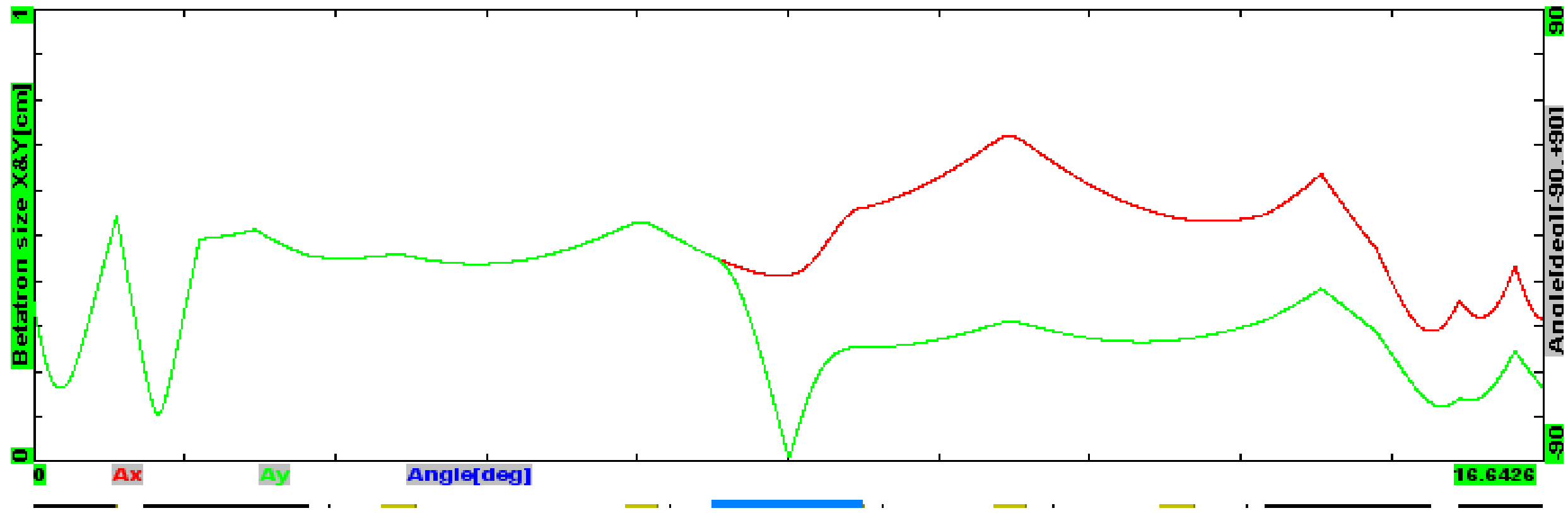

**Figure 3.4.** Beam sizes with space-charge for Set A.

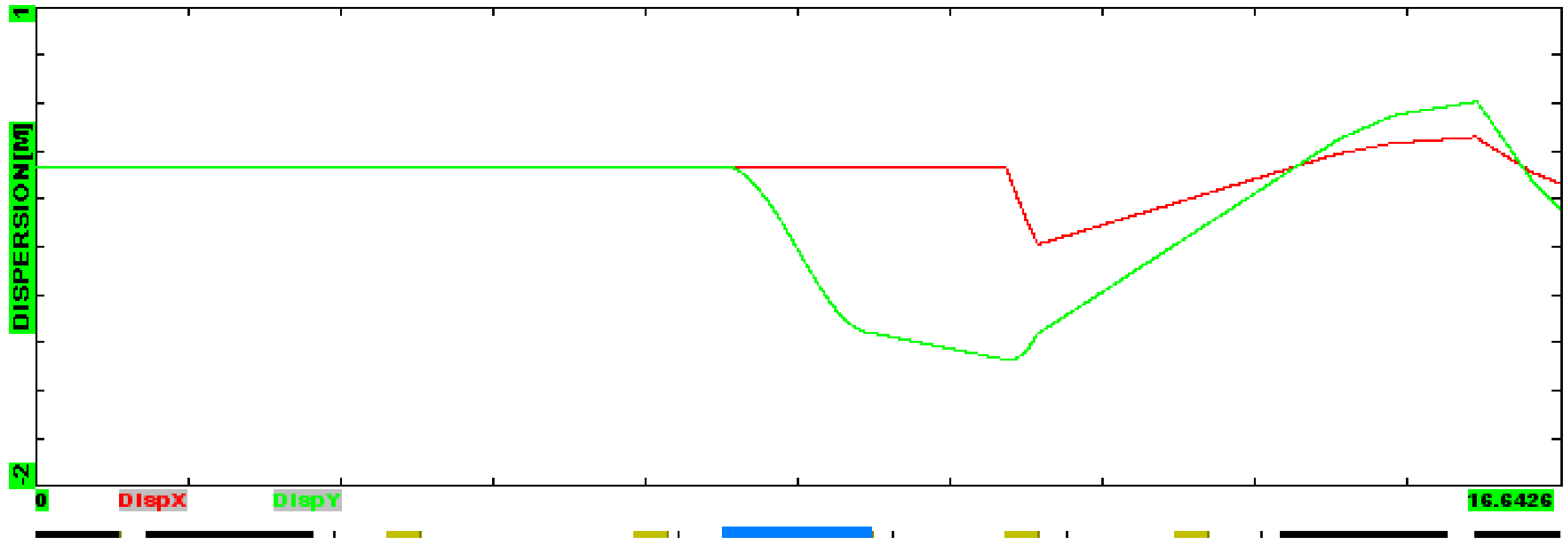

**Figure 3.5.** Dispersion functions for Set A.

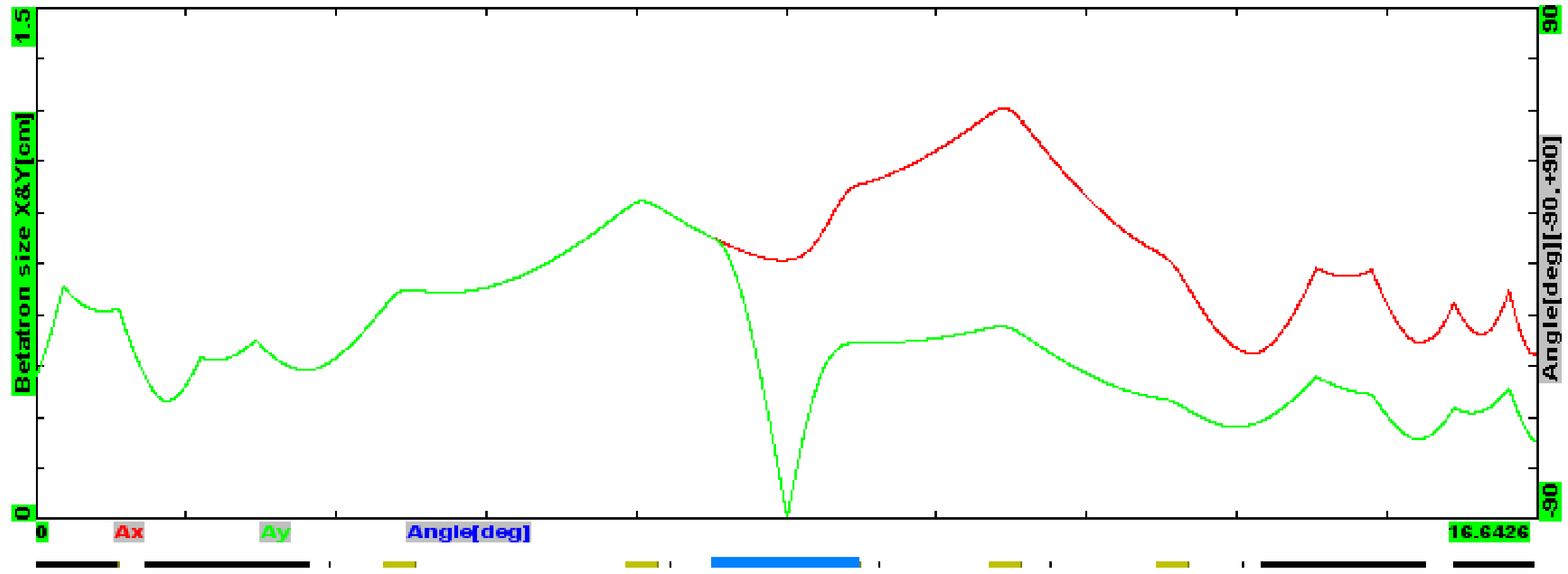

**Figure 3.6.** Beam 'sizes' with space-charge for Set B.

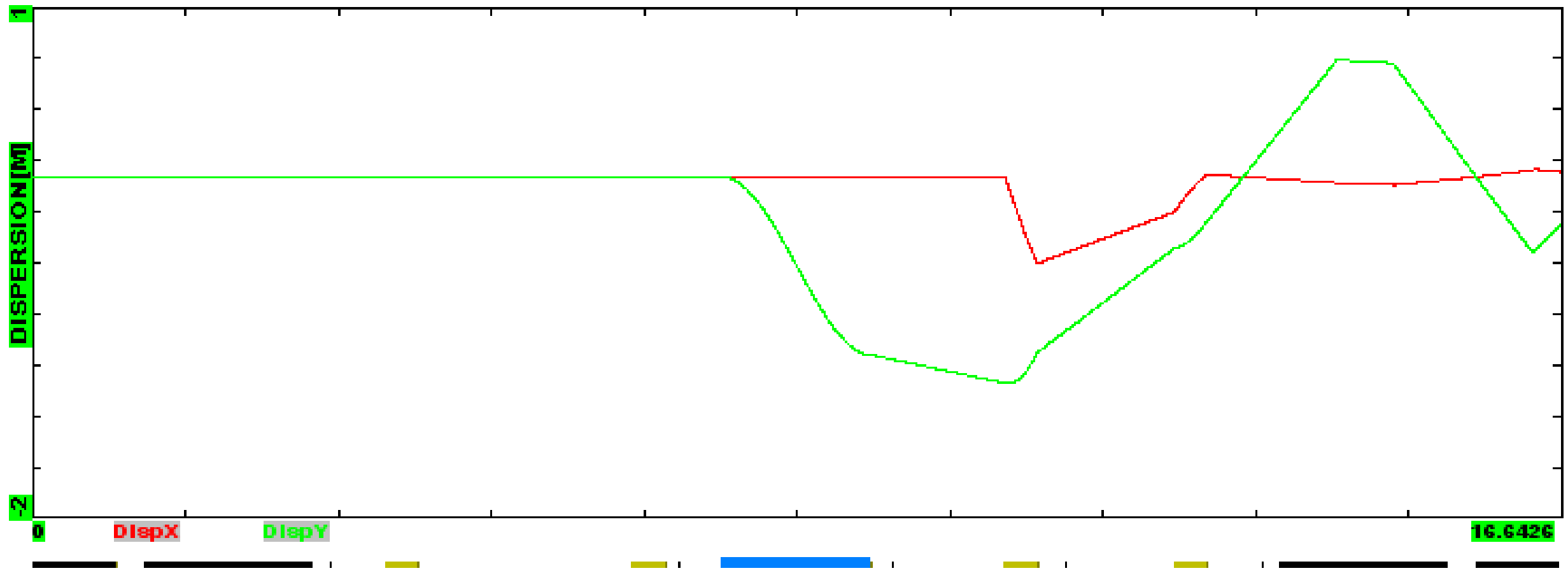

**Figure 3.7.** Dispersion functions for Set B.

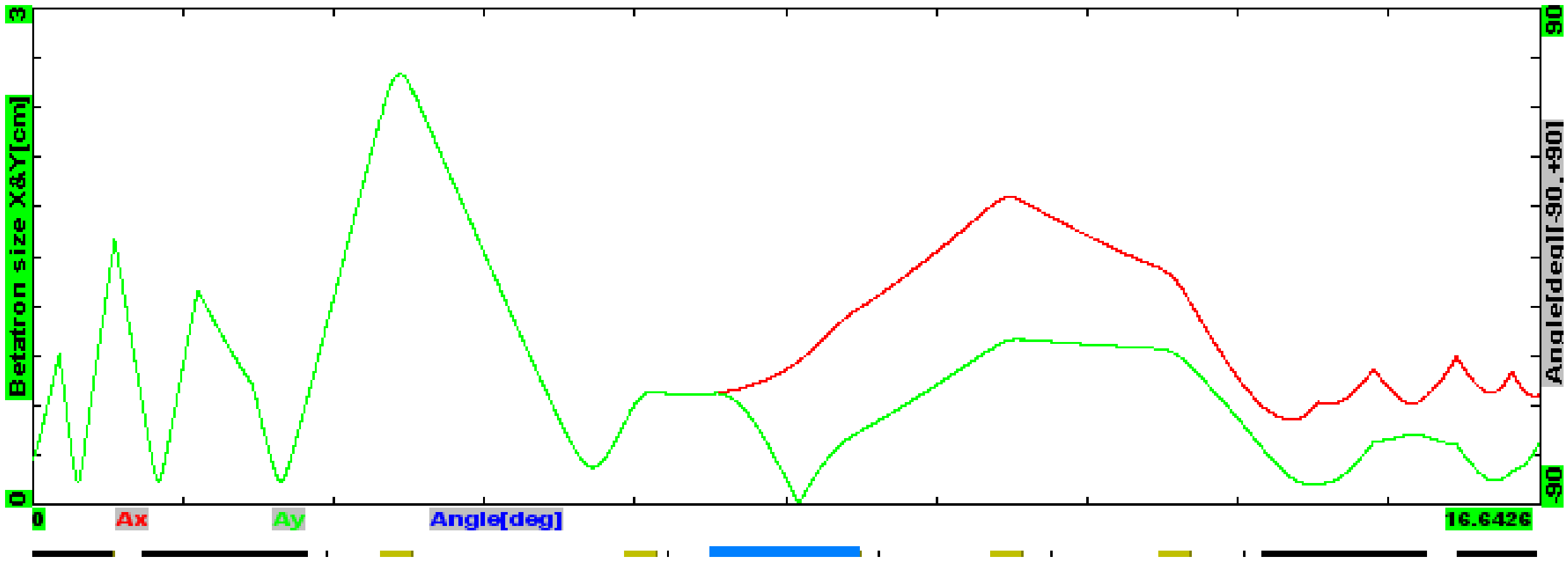

**Figure 3.8.** Beam sizes with space charge for Set C.

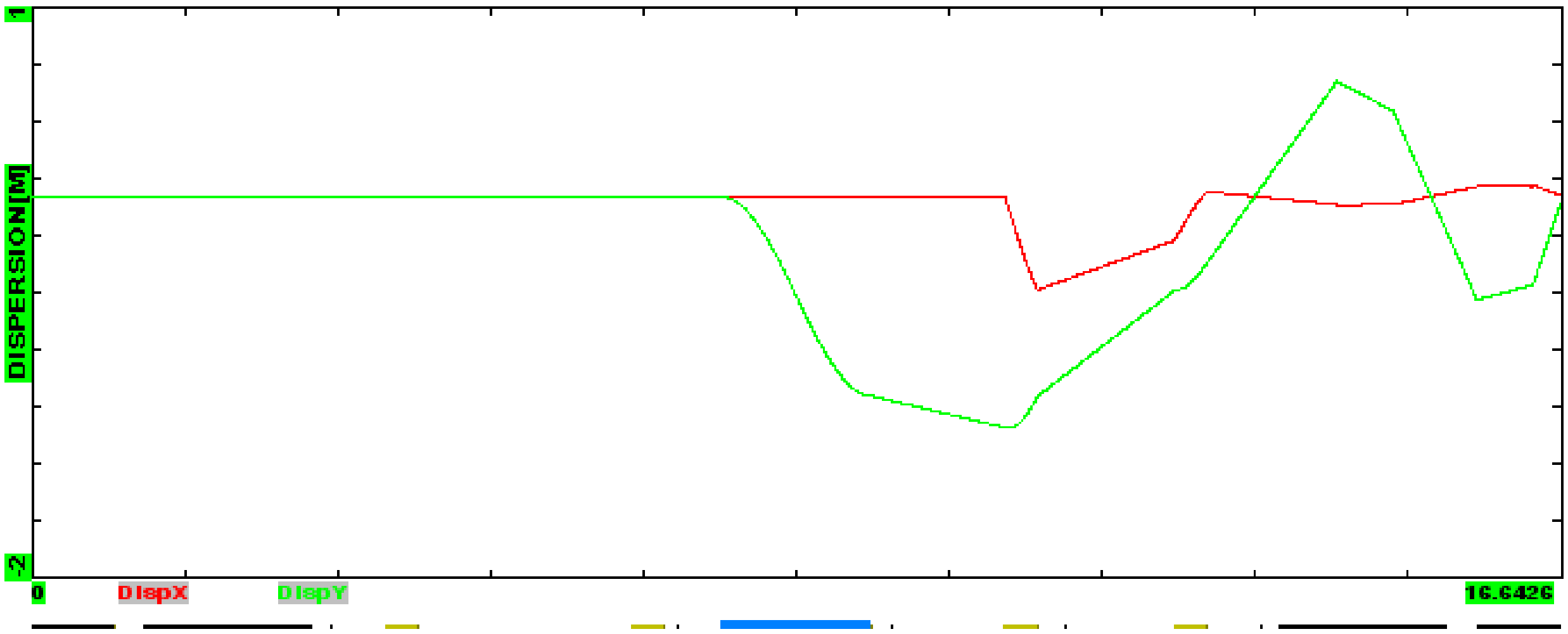

**Figure 3.9.** Dispersion functions for Set C.

Note that the vertical axes are not the same for all the beam sizes plots. In general, we were able to achieve good clearance, in particular in the deceleration column. There are a few areas where the beam becomes fairly large for Set C, but the aperture is large at this location.

The table below summarizes the final focusing settings obtained for all three cases. Note that we do not perform our simulations all the way to the collector entrance thus we do not have settings for the last two focusing elements (SPD01 and SPD00).

**Table 3.1.** Summary of the found settings. The list of the focusing elements is ordered according to the beam direction. All current values are in Amperes.

| | **Set A** | **Set B** | **Set C** |
|---|---|---|---|
| ***SPAGNI*** | 0.6 | 1.8 | 1.8 |
| ***SPA00I*** | 1.5 | 5 | 5 |
| ***SPA01I*** | 1.5 | 2 | 3 |
| ***SPA02I*** | 2.5 | 2.5 | 5 |
| ***SPA03I*** | 4.5 | 3 | 5 |
| ***SPA04I*** | 4 | 3.5 | 5 |
| ***SPA05I*** | 2 | 3.5 | 5 |
| ***SPA06I*** | 3 | 5.5 | 7.7 |
| ***SPA07I*** | 4.5 | 4.5 | 10 |
| ***SPD07I*** | 4.4 | 4.6 | 4.4 |
| ***SPD06I*** | 0 | 5 | 5 |
| ***SPD05I*** | 3 | 3.5 | 4.3 |
| ***SPD04I*** | 2 | 3 | 4 |
| ***SPD03I*** | 3.5 | 3.3 | 3.5 |
| ***SPD02I*** | 3 | 3 | 3 |

To simulate the onset of a full discharge, the anode voltage was reduced from 10 kV to 7.5 kV while the focusing settings were kept unchanged. For both Set A and B, the settings found were deemed quite robust. Below are the screenshots obtained for Set A (similar data exist for Set B).

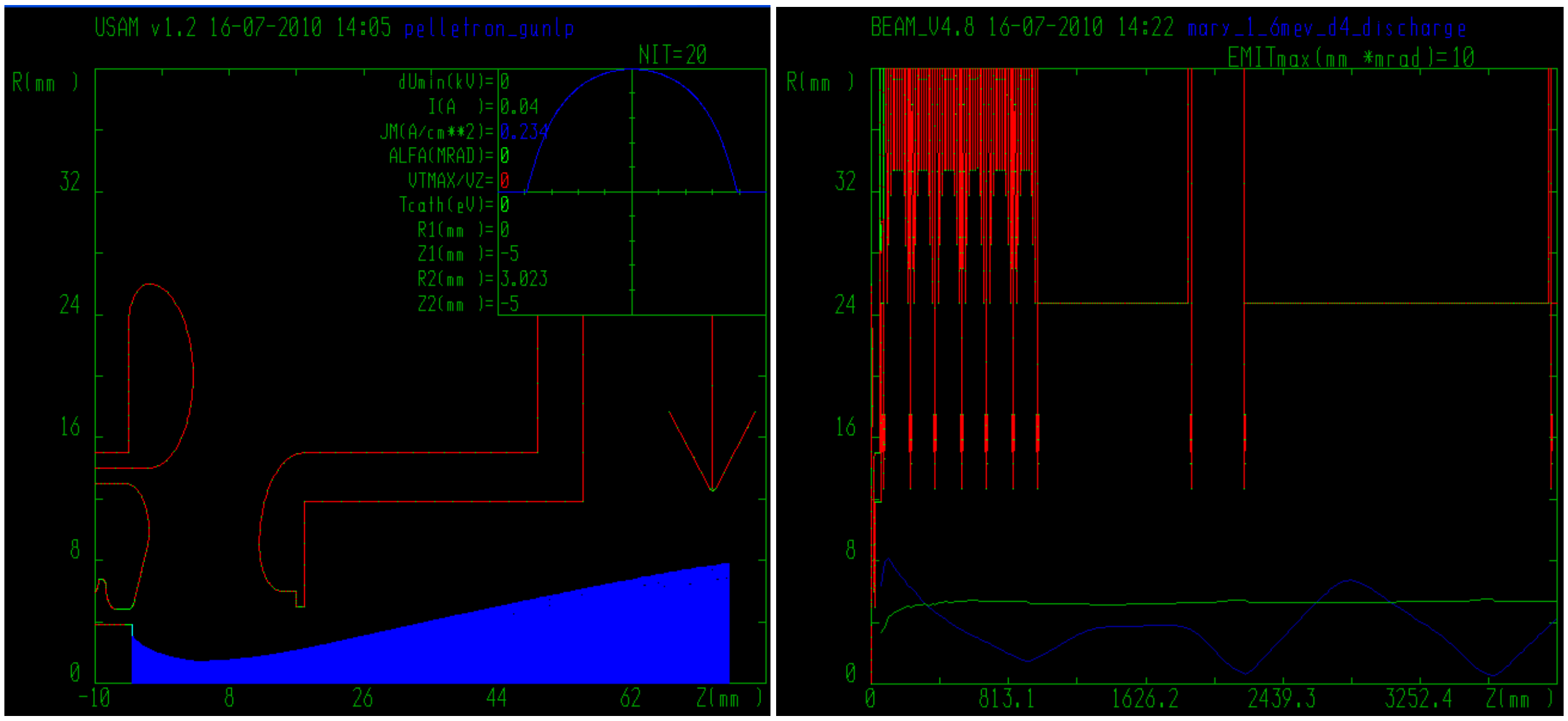

**Figure 3.10.** Gun simulation (left) with the beam current density profile at the cathode in the upper right corner and beam envelope from BEAM (right).

A full discharge 'case' was not simulated for Set C because while starting the studies for cases A and B, we observed that the terminal voltage of 1.6 MV did not pose any full discharge problems. In fact, during the entire study period, whether the beam was pulsed or DC, the Pelletron did not undergo a single full discharge (or even, unprovoked recirculation interruption). This was not at all our experience when commissioning the Pelletron at 4.3 MV in 2005.

# IV. HV performance

The Pelletron High Voltage (HV) regulation system and its performance at nominal operation conditions (4.3 MV) are described in detail in Ref. [4.1]. A simplified schematic is shown in Fig. 4.1.

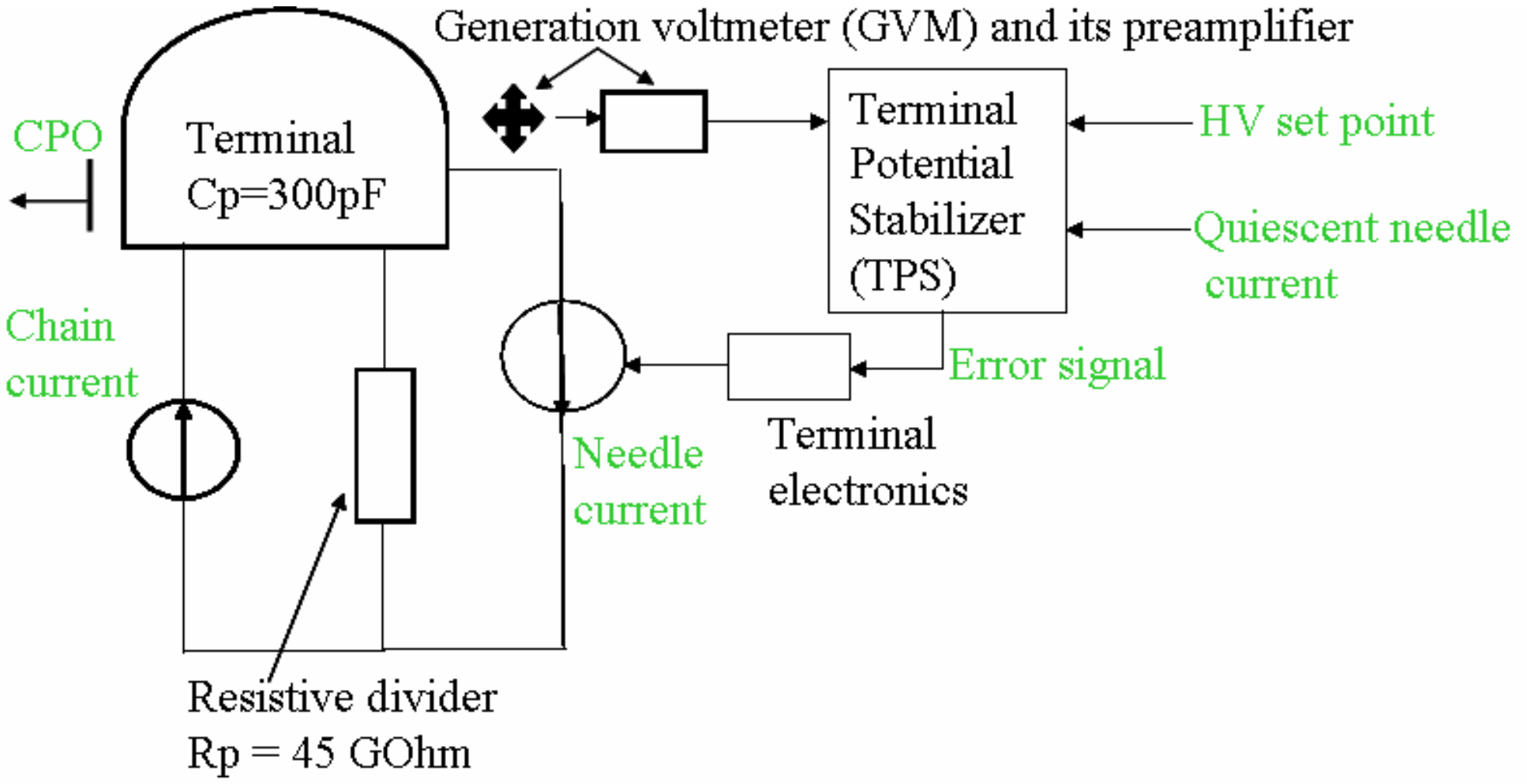

**Figure 4.1.** Pictorial view of the Pelletron energy regulation system.

The Pelletron chain delivers an (ideally) constant current $I_{ch}$ to the HV terminal, which returns to the ground primarily through a set of three parallel resistive dividers. For simplicity, we will describe the set as a single resistor $R_p$ = 45 GOhm so that the Pelletron voltage $U_p$ is approximately $U_p \approx I_{ch} \cdot R_p$.

To compensate the HV ripple caused by fluctuations of the chain current, corona currents through the insulating gas, and the beam loss, a small current $I_n$ is emitted from a set of three needles protruding from the terminal shell toward the tank wall. The terminal voltage is measured by a Generation Voltmeter (GVM), its value is compared with the set point in the analog Terminal Potential Stabilizer (TPS), an error signal is generated taking into account the value of the desired quiescent needle current, and the terminal electronics generates a corresponding needle current. The described loop has a finite maximum gain, which is usually reduced to avoid instabilities using the so-called "control gain" parameter G (R:CTLGAN in ACNET, adjustable from 0 to 100%).

A possible way to describe the performance of the regulation system is to compare the slope $dU_p/dI_{ch}$ with and without regulation. Without regulation, $dU_p/dI_{ch} = R_p$, and with regulation

$$\left(\frac{dU_p}{dI_{ch}}\right)^{-1} = \frac{1}{R_p} + \frac{G}{R_{eff}}, \qquad (4.1)$$

where $R_{eff}$ is the characteristic of the maximum possible gain. For $U_p$ = 4.3 MV, $R_{eff}$ was found to be ~90 MOhm [4.1]. Results of a similar measurement in this study are presented in Fig.4.2.

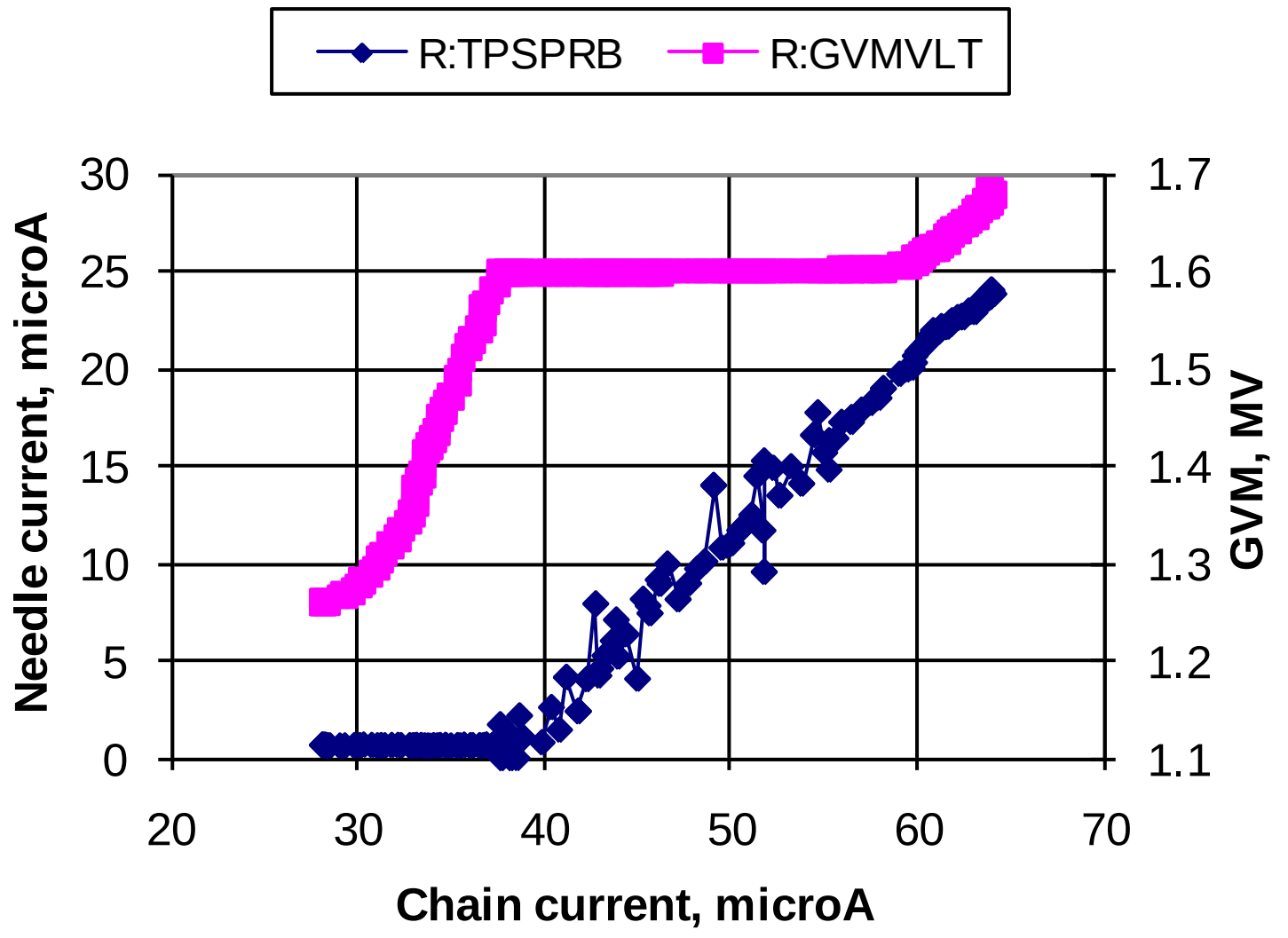

**Figure 4.2.** The terminal voltage (as reported by R:GVMVLT) and the needle current (R:TPSPRB) as a function of the chain current (R:CHN1I). Parameters of the HV regulation system during the measurement were: the HV set point R:TPSTRV = 1.6000 MV; the quiescent current R:TPSPRQ=10 µA; the needle position R:TPSPRD = 25 mm; and the control gain G=R:CTLGAN = 60%.

For a low chain current, the HV is below the set point, and the regulation circuitry suppresses the needle current. As the result the HV increases linearly with the chain current as $U_p = I_{ch} \cdot R_p$. A deviation from linearity between the HV and the chain current at the beginning of the curve is related to a too fast increase of the chain current that resulted in a significant portion of the current to be $C_p \cdot dU_p/dt$ ($C_p$ is the terminal-to ground capacitance). When the HV is close to the set value, the regulation circuitry adjusts the needle current roughly equal to the increase of the chain current. The slope of the HV curve in the region of regulation gives the same value ~90 MOhm as for the case of regulation at 4.3 MV. Therefore, one should expect to have the same suppression of the chain current fluctuations as well.

A part of the terminal electronics in Fig. 4.1 is a triode tube. The needles are connected to the anode of the triode, and the needle current is regulated by adjusting the grid voltage of the tube. The present schematic allows operating the grid voltage only at negative potentials, and it limits the needle currents in the regulated mode. As a result, when the increase of the chain current forces the grid voltage to reach zero, the regulation is disabled, and the HV starts to increase faster again (the area on Fig. 4.2 at $I_{ch}$ > 60 µA. Consequently, the maximum needle current with a good regulation is 15 µA. Note that because of concerns about the protection system, most of the measurements were made at a significantly lower current of 2 µA.

The maximum needle current at a given HV depends on the needle position with respect to the terminal shell. The present safe range of the needle motion is 25 mm. When the parameter that determines this position, R:TPSPRD, is zero, the needles are parked inside the terminal and cannot emit any current; it is used for the Pelletron conditioning. Normal operation position at 4.3 MV is 10-15 mm. For the low-energy operation, the needles were moved to the most outward position, 25 mm to maximize the possible needle current. An attempt to run the Pelletron at even lower HV, 1 MV, showed that at R:TPSPRD = 25 mm the maximum needle current in the regulation mode drops to ~1 µA, which does not allow operation with beam using the present configuration. If operation at 1 MV is needed, modification of either the needle motion system or the triode circuitry will be necessary.

# V. Tools and methodology for tuning

In this section a brief description of the tuning procedure which was used during the measurements is summarized. First, we introduce the tools/diagnostics available.

## *A. Beam Position Monitors (BPM)*

There are 5 BPMs sets in the beam line (each set measures the x and y positions), all outside of the acceleration and deceleration columns. They are designed to work in both pulsed and DC mode. In addition to the beam offsets (trajectories), the monitors return an intensity proportional to the beam current. However, if electrons hit a BPM, the intensity read-back increases due to the plates collecting additional charges. By looking at the different BPM intensities along the beam line, one can identify where the beam might scrape simply by the enhanced BPM intensity near that location.

## *B. Anode power supply current (also called bias current)*

Nominally, the bias current is proportional to the anode voltage. However, when losses occur, the anode current increases compensating for the beam loss (*i.e.* current not returning to the collector). In turn, this behavior is used to assess how well the beam is transported to the collector.

## C. Radiation Monitors (also referred as loss monitors)

There are a number (4) of gas-filled radiation monitors ('BNL type') along the U-bend beam line, all outside the Pelletron. They are used by the Pelletron Protection System to abort the beam when losses are higher than a certain threshold. In this study the monitors were less useful than in the standard 4.3 MV mode because the signal of the monitors at 1.6 MV was significantly lower for the same amount of the current loss.

## D. Capacitive Pick Out (CPO)

The Capacitive Pick Out measures the change of the terminal voltage and its output can be displayed on an oscilloscope. In pulsed mode, this signal can be used to identify the level of the total losses: with no losses the CPO drop is determined by the amount of charge leaving the terminal to fill the beam line. Because the total flight time through the beam line, ~0.1 µs << the pulse length, ~2 µs, the CPO waveform roughly mirrors the beam current monitor waveform. With losses the terminal becomes more positive, *i.e.* the CPO signal steps up and returns to its DC level on a much longer time scale than the few microseconds that the pulse lasts.

## E. Tube Monitors (TM)

The tube monitors measure the AC component of the current that flows through the last (bottom) electrode of each tube. In the pulsed mode these signals are displayed together with the CPO signal on the oscilloscope. If the CPO signal indicates losses, the tube monitor signals can give an indication of where the loss occurs. Several scenarios are possible:

i. Both signals have the same DC offsets before and after the pulse (Fig.5.1). In this case, the signals are proportional to the beam current flying through the electrode because of capacitive coupling. In part, if the Deceleration TM signal is lower than Acceleration TM's, a significant portion of the beam is lost in the beam line.

ii. Offsets of the Acceleration and Deceleration TM signals before and after the pulse are different (Fig. 5.2). It indicates a beam loss coming to some of electrodes in the tubes. Two tubes are electrically connected by separation boxes. Therefore, if the entire loss occurs above the first separation box, both TM signals are similar. If the final offset of the Deceleration TM is larger Acceleration TM in absolute value, the beam loss comes primarily to the bottom electrode of the deceleration tube.

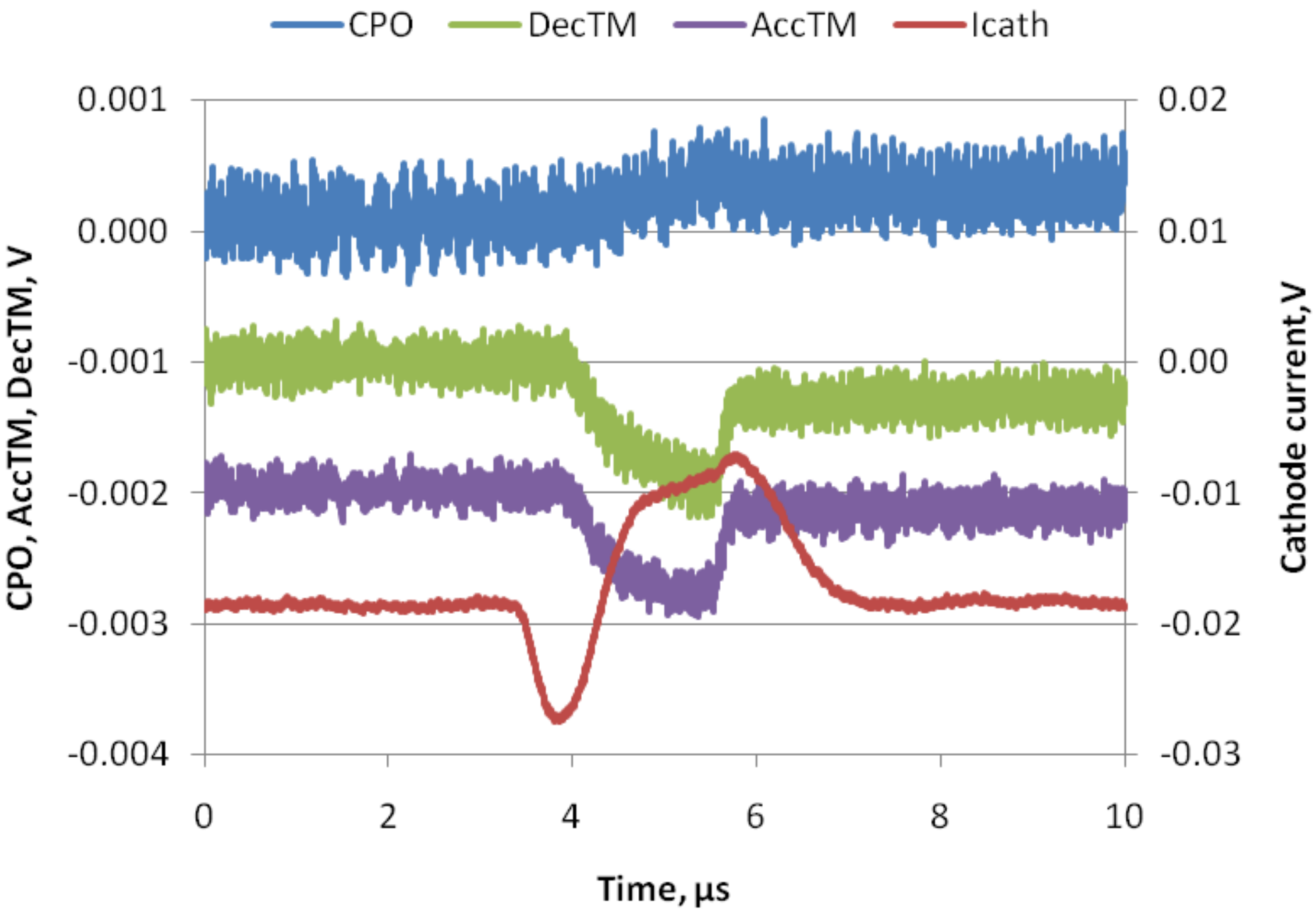

**Figure 5.1.** Example of raw oscilloscope traces of the CPO signal, the tube monitors and the cathode current for a case of low beam losses. 12-Aug-10. All signals are in mV.

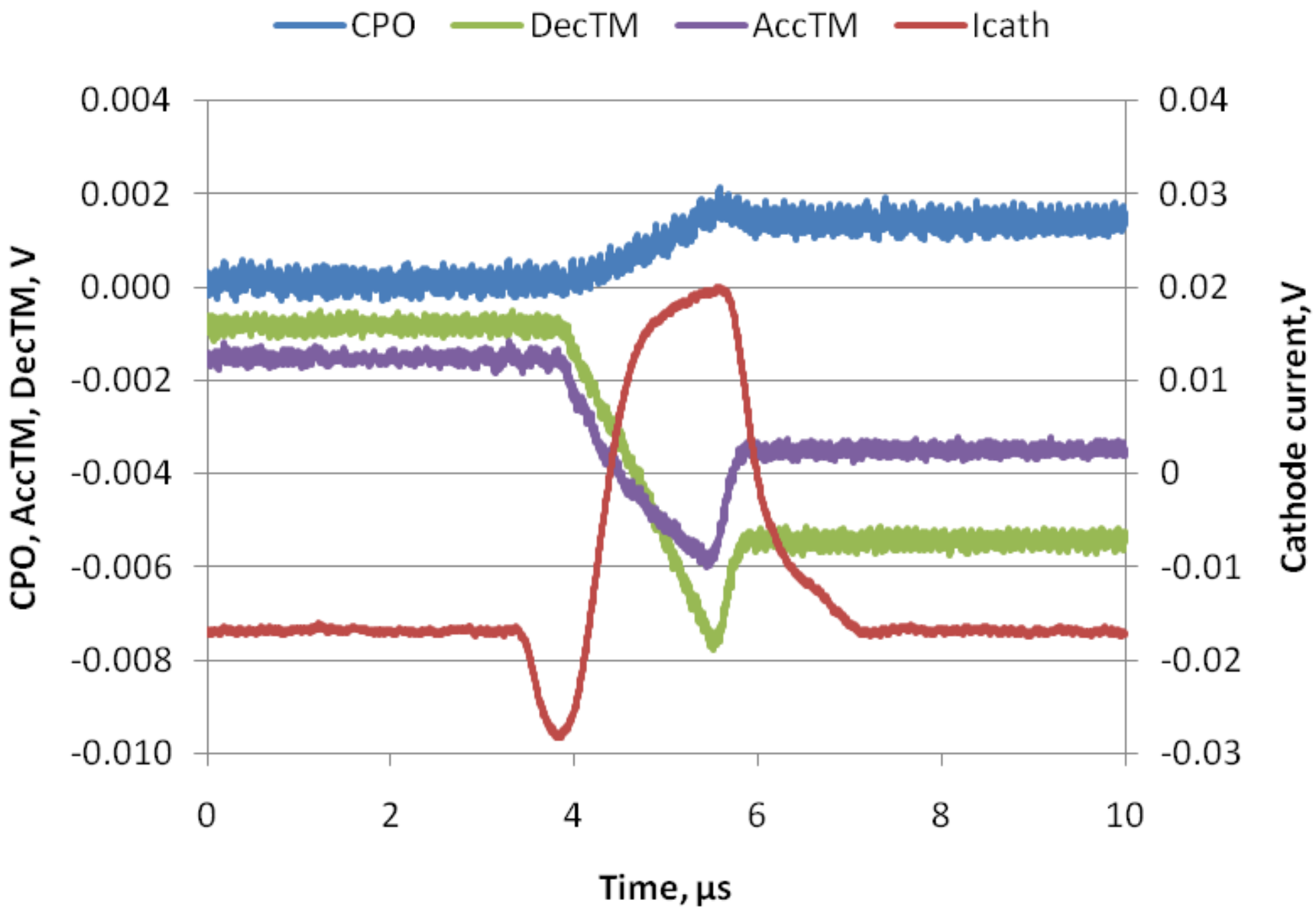

**Figure 5.2.** Example of raw oscilloscope traces of the CPO signal, the tube monitors and the cathode current for the case of a beam loss inside the deceleration tube. 12-Aug-10. All signals are in mV.

## F. Tuning procedure

The charging power supplies voltage is set to ~6 kV which produces an accelerating voltage of 1.6 MV. The anode voltage is set to the desired value (10 or 20 kV in these studies).

The first step is to operate in the pulsed mode and pass the beam into the collector cleanly *i.e.* no beam loss. For the anode voltage of 10 kV, a DC voltage of -2.5 kV on the control electrode, $U_{ce}$, maintains the gun closed (*i.e.* no emission from the cathode). At 20 kV on the anode, $U_{ce}$ = -5 kV to keep the gun closed. To pulse the beam, the control electrode is made more positive (or less negative). For instance, for a pulse amplitude of 1.5 kV, $U_{ce}$ = -2.5kV + 1.5kV = -1 kV during the pulse. The beam pulse width and frequency can be adjusted. Typically, a pulse width of 2 μs at 2 Hz is used.

Initially the Pelletron solenoids are set to the values produced by the simulations with SAM and OptiM, and the dipole correctors are set to zero. Starting with a relatively low pulse amplitude, the beam is tentatively threaded through the beam line. Mostly using the BPM intensities as an indication of losses at first, the orbits at the exit of the acceleration tube (first two BPMs) are adjusted so that, losses there are minimal and a BPM signal is observed right after the bend (Bx/yD08 in our current nomenclature). Typically, one uses the dipole correctors directly upstream of the BPM considered for tuning the corresponding part of the beam line. To progress down the beam line, scans of the BPM intensity vs. corrector setting are done and correctors set to the middle of the window with no losses. In the deceleration column, since there are no BPMs, only the bias current and tube monitors can be used to eventually reach the collector cleanly. Figure 5.3 illustrates the scanning process. On the plots, the intensities for two successive BPMs are plotted against the current of the corrector located just upstream of the first of the two BPMs.

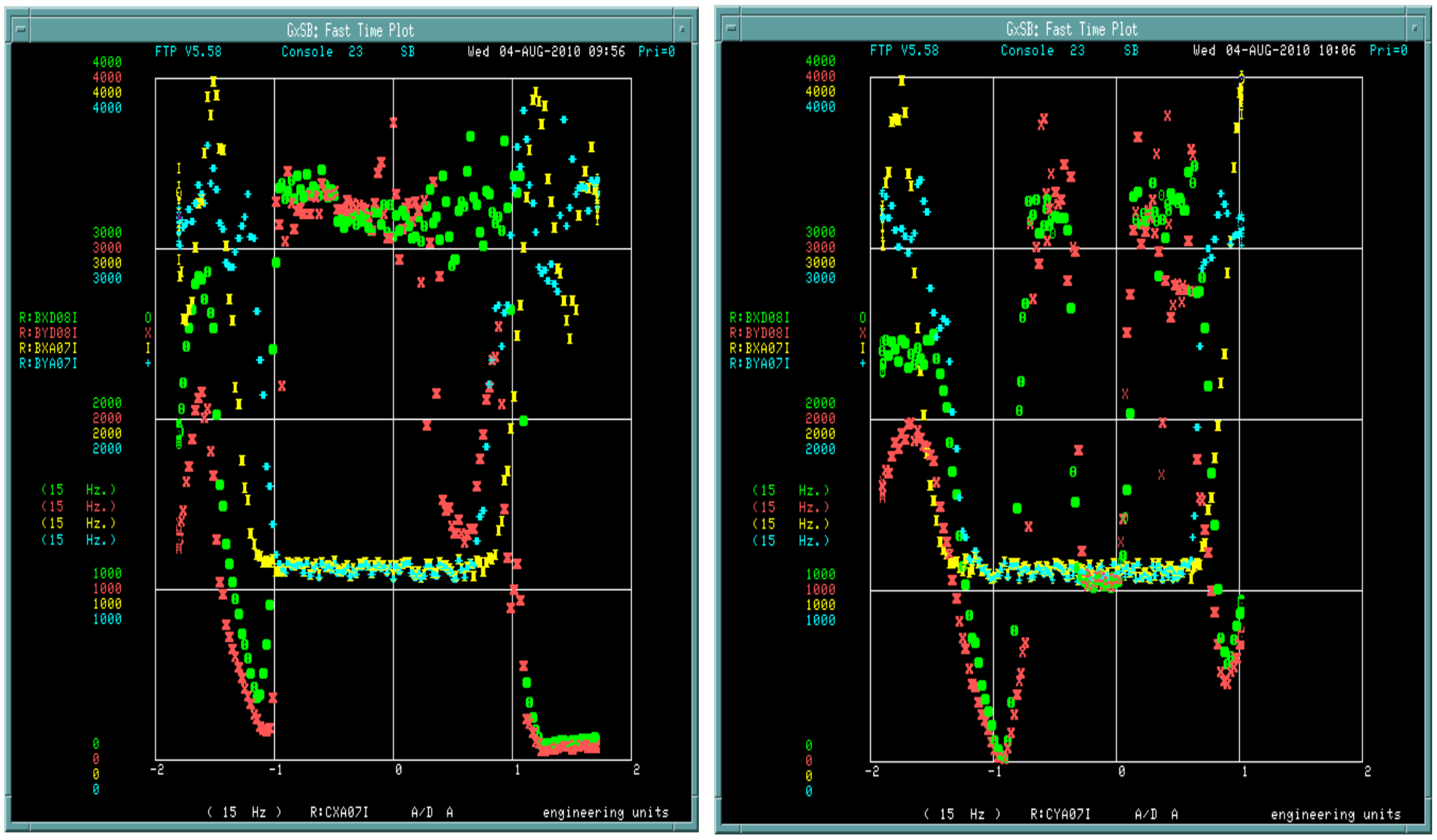

**Figure 5.3.** Scan of the D08 BPM with the A07 corrector. D08 is downstream of the A07 BPM and correctors. Left: scan in the x-direction, which shows no settings with no loss; Right: scan in the y-direction, which shows a narrow window with no loss.

The very different intensities in the A07 BPM ($1^{st}$) and the D08 BPM ($2^{nd}$) indicate that the beam scrapes somewhere between the two. On the left plot, the scan shows no good region. On the right plot, one can see a very small region where the A07 and D08 BPM intensities are equal, indicating that the beam is going through the D08 BPM over that range of corrector currents.

In the next step the pulse amplitude is increased. Depending on whether or not the BPM intensities remain all equal, which means still no loss in the beam line, the focus remains on orbit correction or envelope correction (starting to adjust solenoid lenses). If the beam cleanly reaches the deceleration column, the Tube Monitors and the CPO signals on the oscilloscope are the diagnostics to use as described above. Note that in our particular case, the solenoid SPA04 is the one which had the biggest impact on the losses.

Finally, the system was switched to DC operation. In this case, the control electrode is not pulsed and $U_{ce}$ is increased slowly so that beam can be extracted from the cathode. In addition, $U_{ce}$ is modulated at 32 kHz so that the BPMs can report positions. Typically, even though a large amplitude beam pulse can cleanly be transported to the collector, the settings found do not allow the transport of a high current DC beam. However, it is usually possible to transport a small amount of current cleanly to the collectors with the settings found in pulsed mode and to continue tuning using the bias current as the main diagnostics.

To do so, two live plots are used. First, the bias current (BIASI) is plotted against the beam current (COLLI) (Figure 5.5). If the beam does not scrape, the bias current is linear with the beam current. Then, as the beam current increases, the beam starts to scrape, and the bias current grows nonlinearly. In the second plot BIASI is plotted against the time.

At this stage, tuning starts by setting the beam current so that to the bias current is the onset of the nonlinear growth. By varying all solenoids and dipole correctors (in small steps), the bias current is returned to its non-scraping value indicating that the beam does not scrape any longer. The beam current is then increased again until the non-linear behavior of BIASI reappears, and the procedure is repeated until the desired beam current is reached and stable.

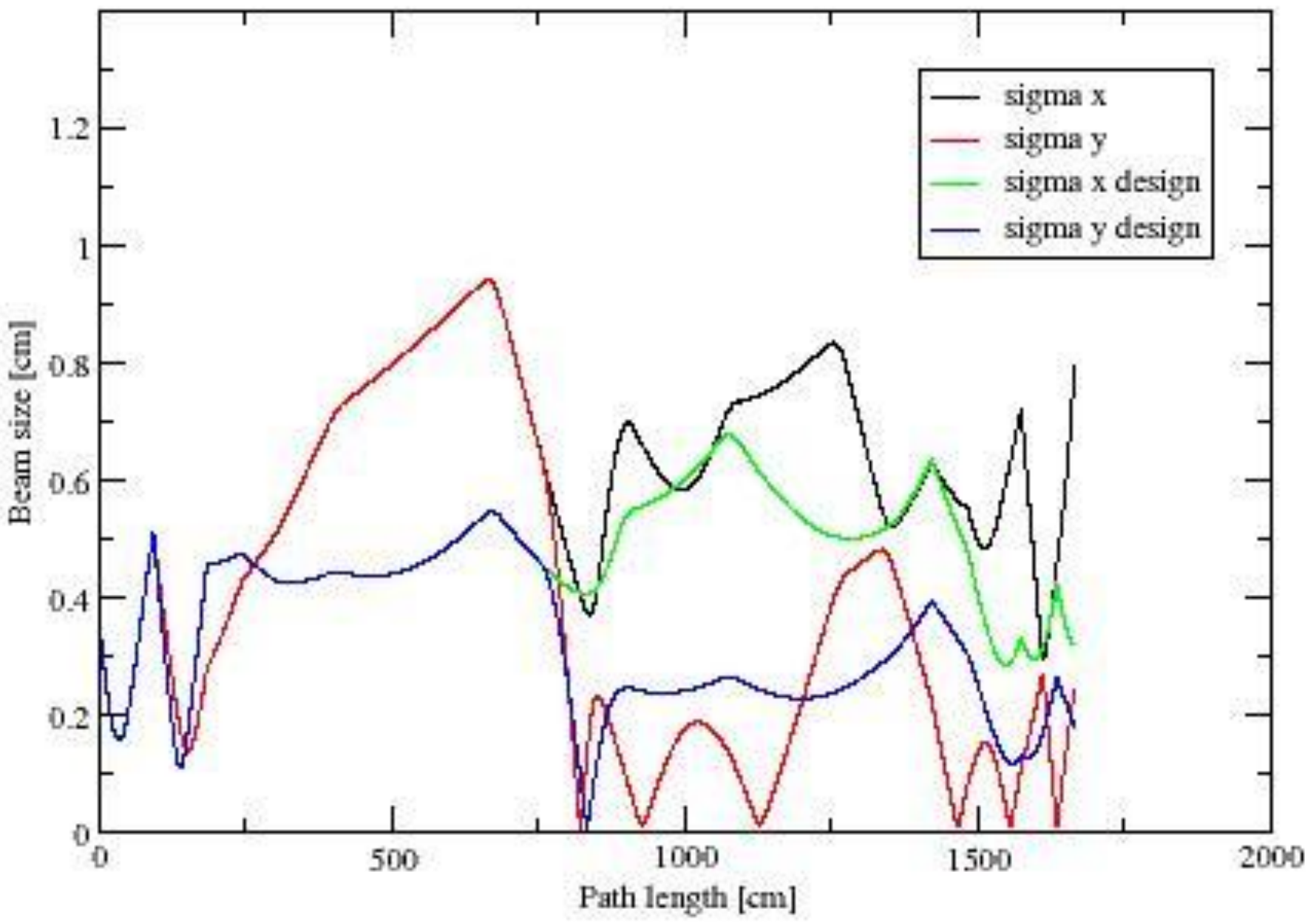

**Figure 5.4.** Simulations of beam envelopes: design vs. experimentally found settings.

As an exercise, for the low magnetic field on the cathode case, the final settings found in the experiment have been put back into the OptiM program and compared to the design settings.

Results are summarized in Fig. 5.4. It seems that we have applied to much change to the focusing solenoids in the accelerating tube, but recovered the beam size after the U-bend.

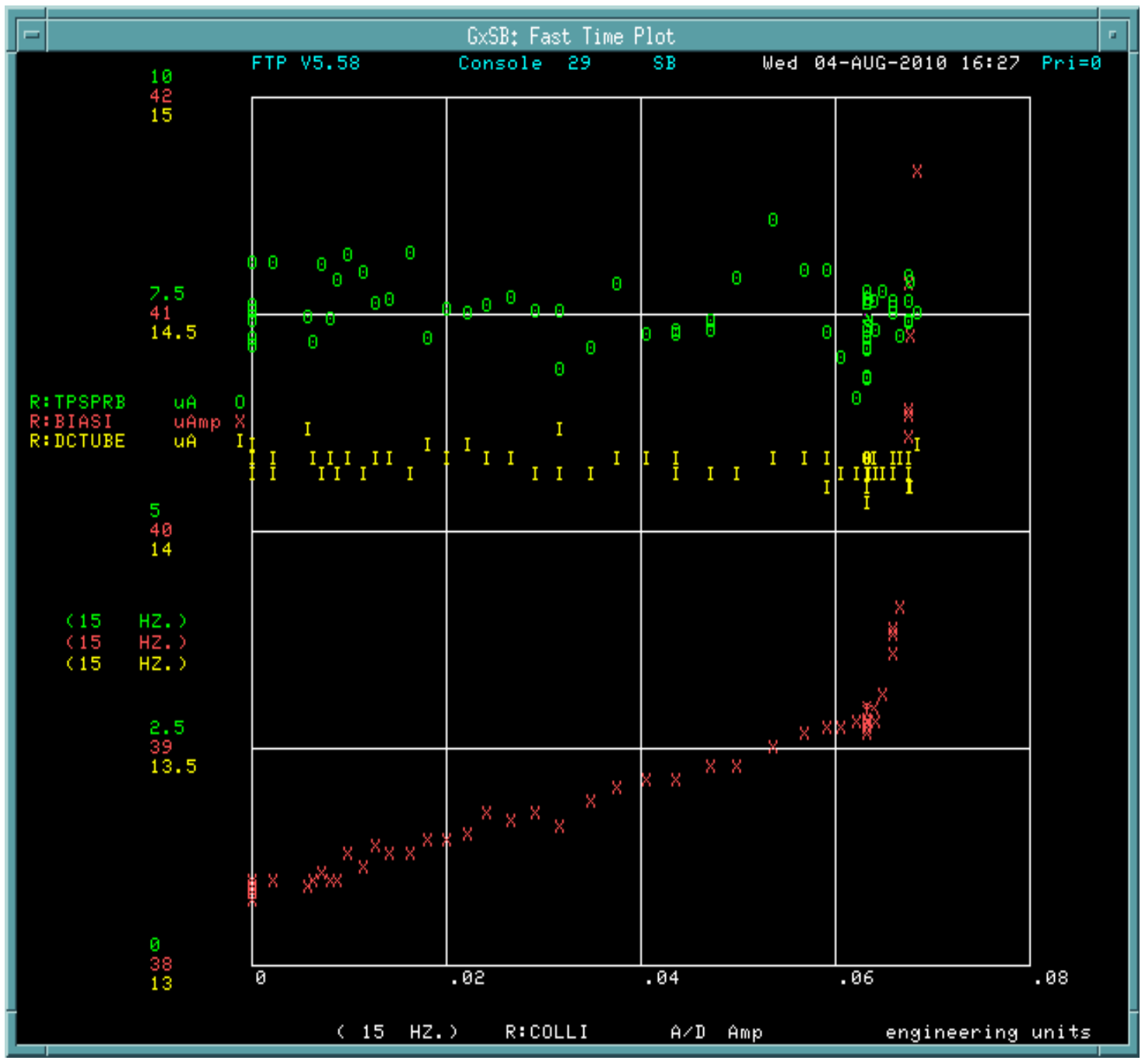

**Figure 5.5.** Scan of the bias current (losses) vs. beam current.

# VI. Results and issues

## *A. Maximum beam current*

The values of the maximum DC current achieved in various configurations during the 1.6 MV run are summarized in Table 6.1.

**Table 6.1.**

| # | Cathode field, G | Anode voltage, kV | Maximum DC current, A | Limit |
|---|---|---|---|---|
| A | 84 | 10 | 0.24 | Maximum gun current |
| B | 253 | 10 | 0.20 | Decision to stop tuning |
| C | 255 | 20 | 0.38 | Decision to stop tuning |

In Case A, the maximum current was limited by a sharp growth of beam losses that did not depend on focusing when the control electrode approached 0 kV. It was interpreted as an onset of the emission from the cylindrical surface of the cathode. According to simulations, at this current the zero equipotential surface (the surface with the potential equal to the cathode's) reaches the edge of the cathode flat surface and goes beyond, which in turn allows extraction of electrons from the cylindrical surface. Trajectories of these electrons differ dramatically from the main beam, and the electrons are lost. This effect determines the maximum gun current at a given anode voltage.

In Case B, the process of increasing the beam current was stopped at 0.2 A, which was the value initially agreed upon. In addition, for the same reason as described above, it is understood that further tuning would not increase the beam current by more than 20%.

Tuning in Cases A and B went faster than it had been planned, and we decided to try operation at a higher current in order to prove that there is no immediate “hard” limit of the maximum current that can be recirculated at 1.6 MV if the anode voltage is increased (Case C). Reaching a beam current of 0.38 A was deemed a proof of this statement.

## B. Recirculation stability

Stability of the beam recirculation was very good in comparison with operation at 4.3 MV. There was not a single full discharge and, moreover, not a single unprovoked beam interruption. All interruptions were related to tuning, and the signals captured by the Transient Recorder looked nearly identical for all of them (Fig. 6.1): the terminal voltage slowly decreases (becomes more positive) because of losses caused by tuning; the cathode current stays almost constant until the protection system detects the decrease of HV and turns the beam off; the Tube Monitors exhibit negative signals indicating a beam loss in the middle of the deceleration tube.

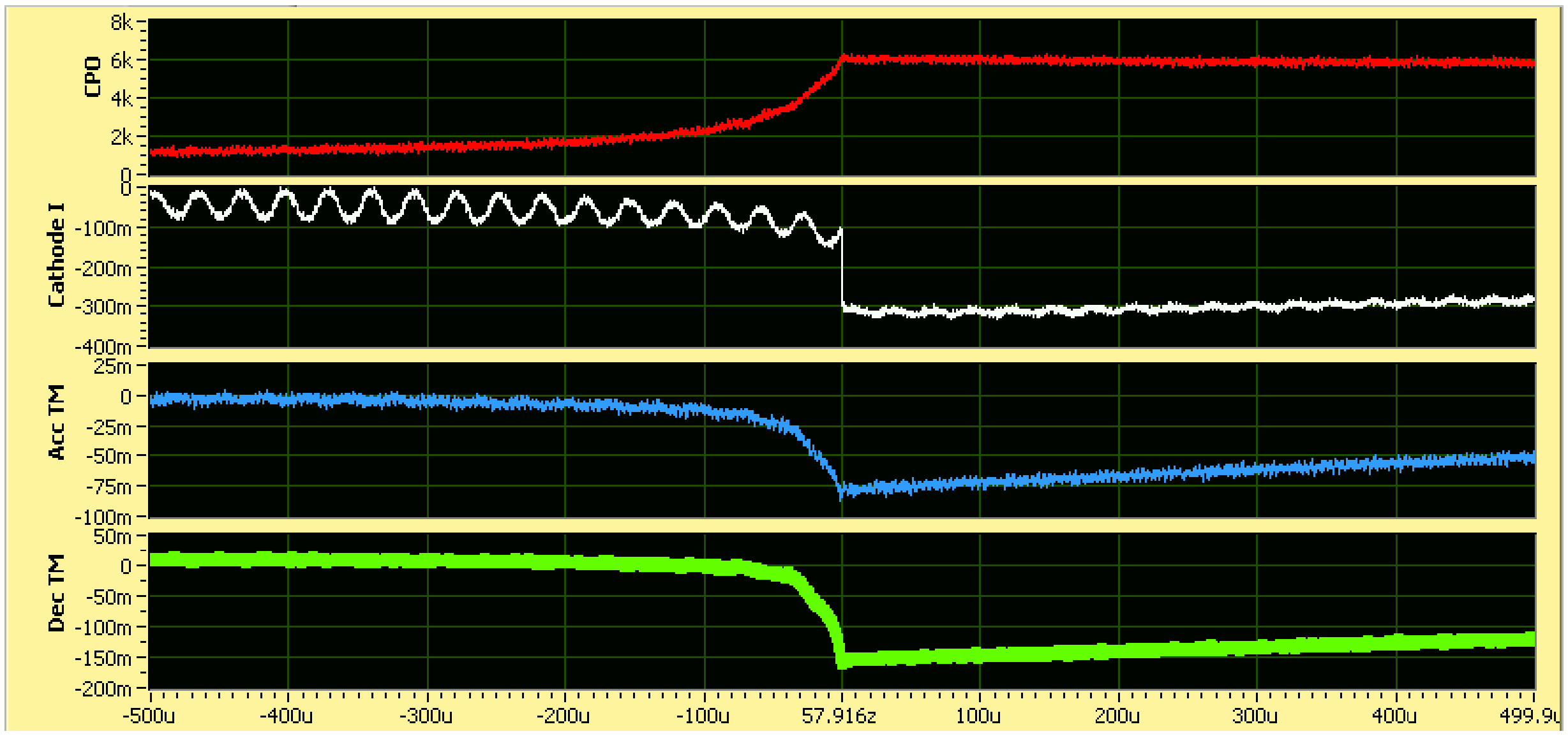

**Figure 6.1.** The Transient Recorder data of an interruption on August 12, 2010, 4:21:12 PM. Settings of the Case #3. The curves, from top to bottom, are: CPO showing the change in the terminal voltage, in KV; the cathode current, in A (AC coupled, the calibration is slightly off); Acceleration and Deceleration tube monitors (raw signals in mV). The DC beam current before the interruption was ~0.38A.

As soon as the beam was left running with the settings that correspond to the linear regime for current losses, it stayed without interruptions. A dedicated run with the settings from Case B and 0.1A current lasted uninterrupted for 20 hours and was stopped intentionally by an operator.

### C. Energy ripple

As it was discussed in section 4, the measured parameters of the regulation loop were found to be the same as for 4.3 MV. The chain current, which fluctuation is the main source of the energy ripple, went down to roughly the same proportion as the HV. Therefore, one should expect to have the relative HV ripple to be independent of the HV, and the value of ~100 V for the ripple found in [4.1] gives a ~40 V ripple (sigma) at 1.6 MV. A beam- based estimation of the energy ripple can be done from spectra of BPM signals (Fig. 6.2). FFT amplitudes peak at several frequencies in all BPMs, but the difference between low- and high-dispersion BPMs is pronounced only at lower frequencies.

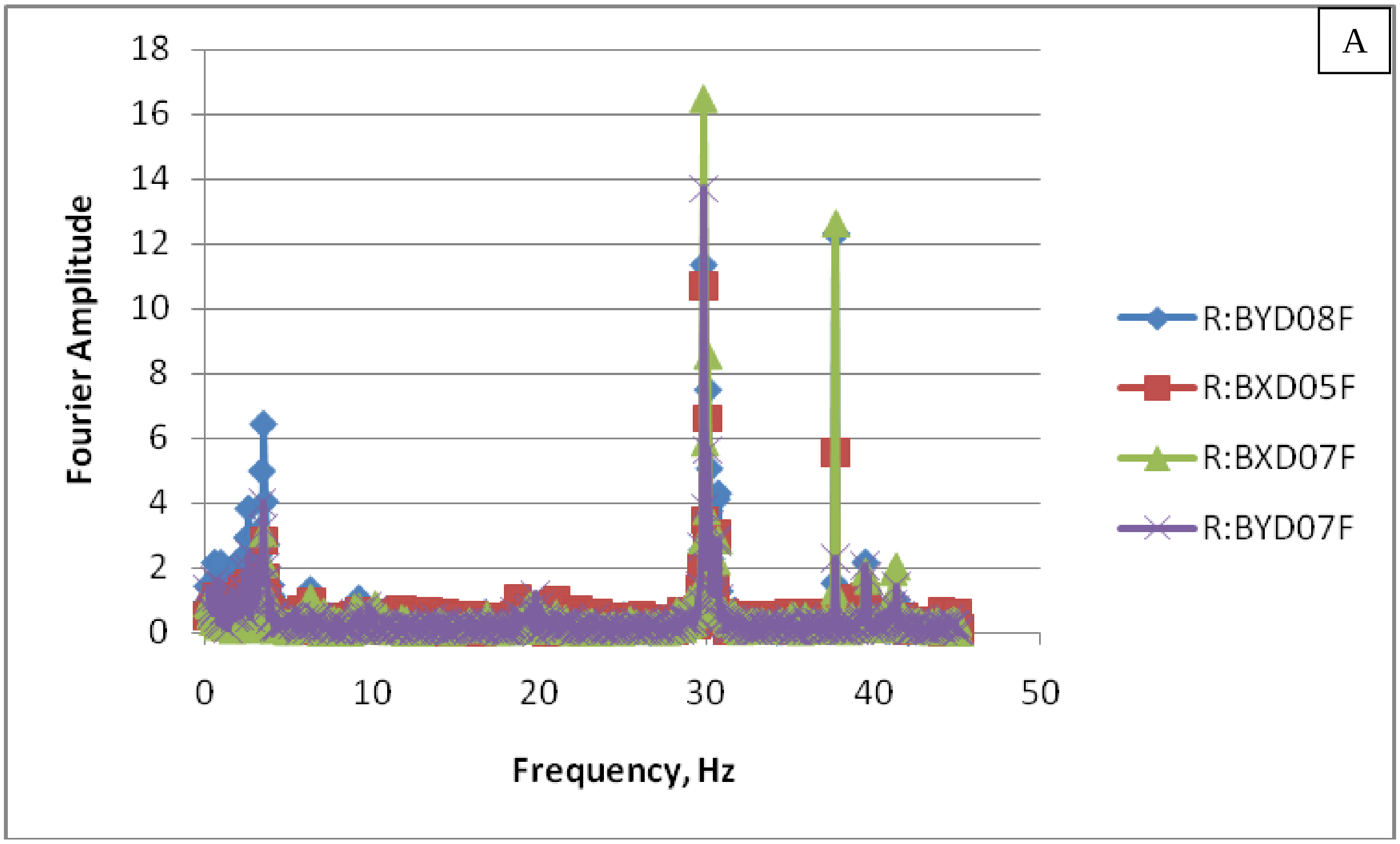

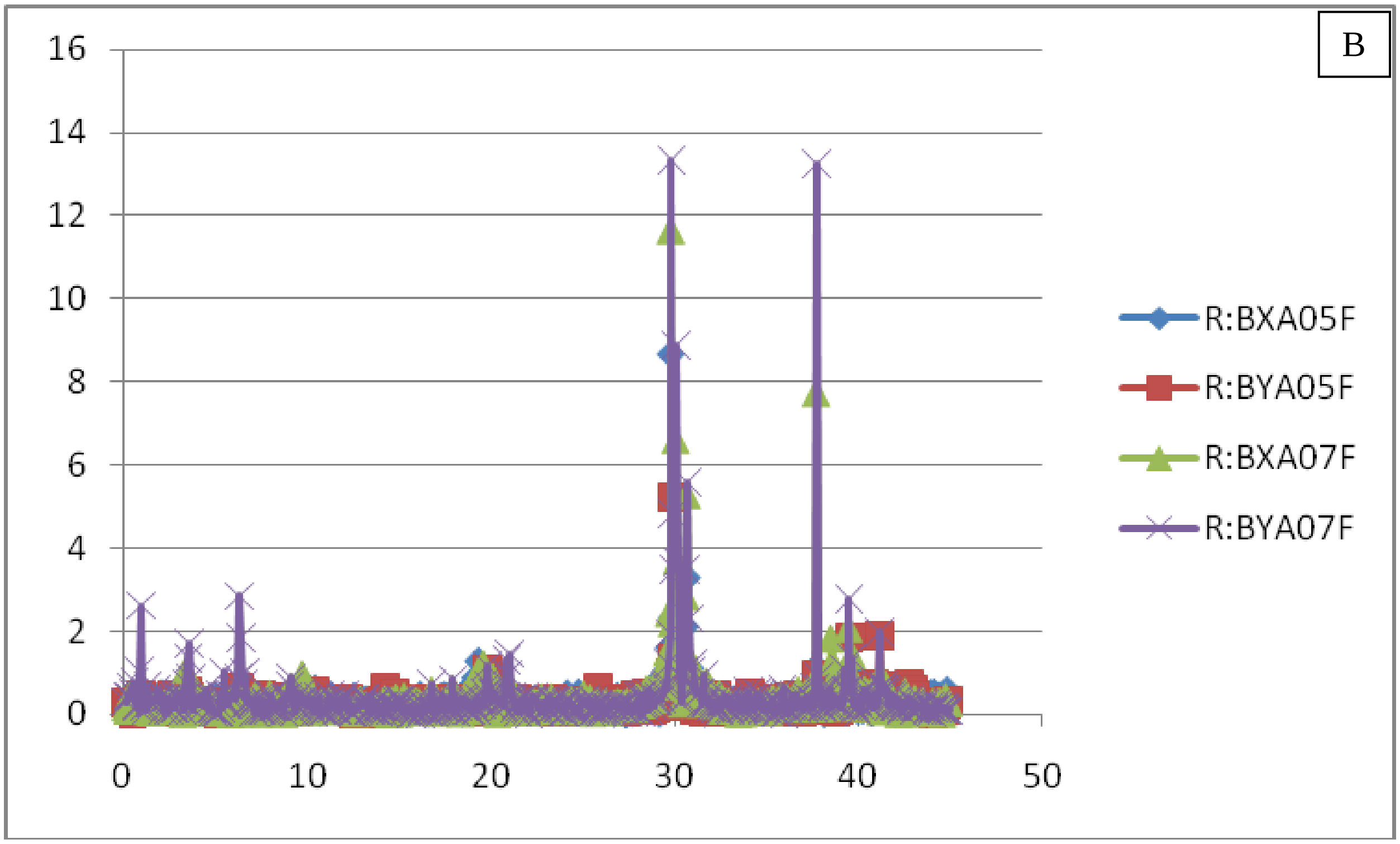

**Figure 6.2**. FFT spectra of 8 BPMs recorded on August 10, 2010. The acquisition frequency was 81 Hz. For FFT in each channel, 1024 recorded points were used. The beam current was 100 mA, Case B. The upper plot (A) presents the BPMs in high-dispersion locations (after the bend), and the lower plot (B) is for low-dispersion locations (before the bend).

By recording trajectories at two different energies, one can reconstruct the dispersion along the beam line. Results of such measurements are shown in Fig. 6.4 as ratios of beam shifts in BPMs to the energy difference.

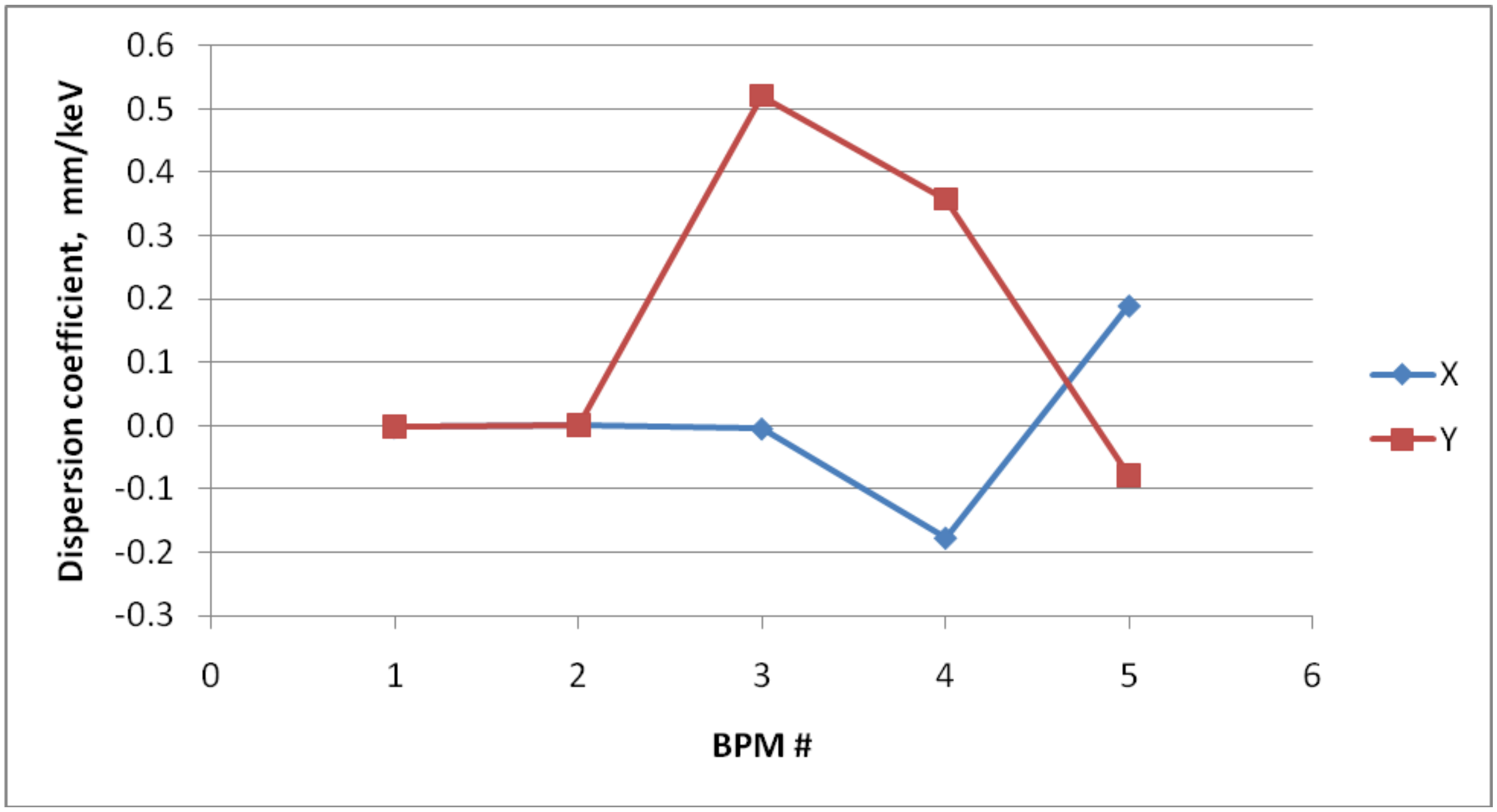

**Figure 6.3.** Dispersion-related coefficients in the beam line. The coefficients were measured by comparing two trajectories with the beam energies differed by 20.7 keV. Settings are of Case B. Beam current was ~1mA, modulation amplitude 200V, high gain. BPM numbering in the plot: 1- A05, 2- A07, 3- D08, 4 – D07, 5 – D05.

BPM signals presented in Fig. 6.2A were normalized by the coefficients shown in Fig. 6.3. For higher frequencies the scatter of FFT amplitudes between these four BPMs did not decrease, but in the range < 5 Hz the amplitudes became close to each other, indicating that the main contribution came from the energy ripple (Fig. 6.4). A 0.5 – 6 Hz filter and the inverse FFT were applied to the data of BPM with the highest dispersion (BYD08). Assuming that this entire signal comes from the energy ripple, the ripple rms is estimated to be 38 eV, in reasonable agreement with the assumption of a constant relative HV ripple.

An upper estimate for the ripple can be done assuming that the all signal of BYD08 in the bandwidth 0.2 – 175 Hz (Fig. 6.5B) is caused by HV fluctuations. This, certainly overestimated value, was found to be 83 eV.

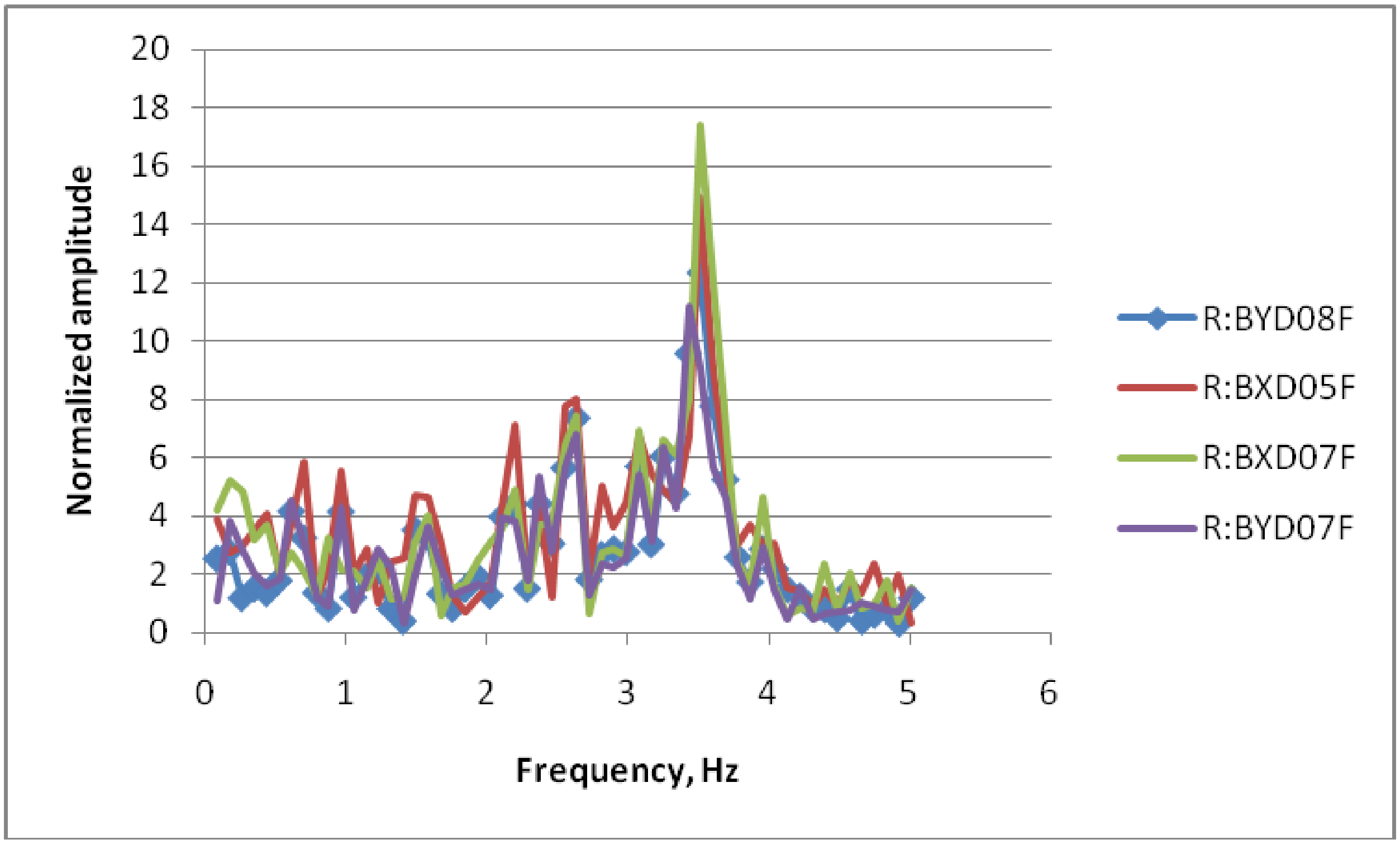

**Figure 6.4.** The same data as shown in Fig. 6.2A but each data set is normalized by a corresponding dispersion coefficient from Fig. 6.3. Only the low frequency portion is presented.

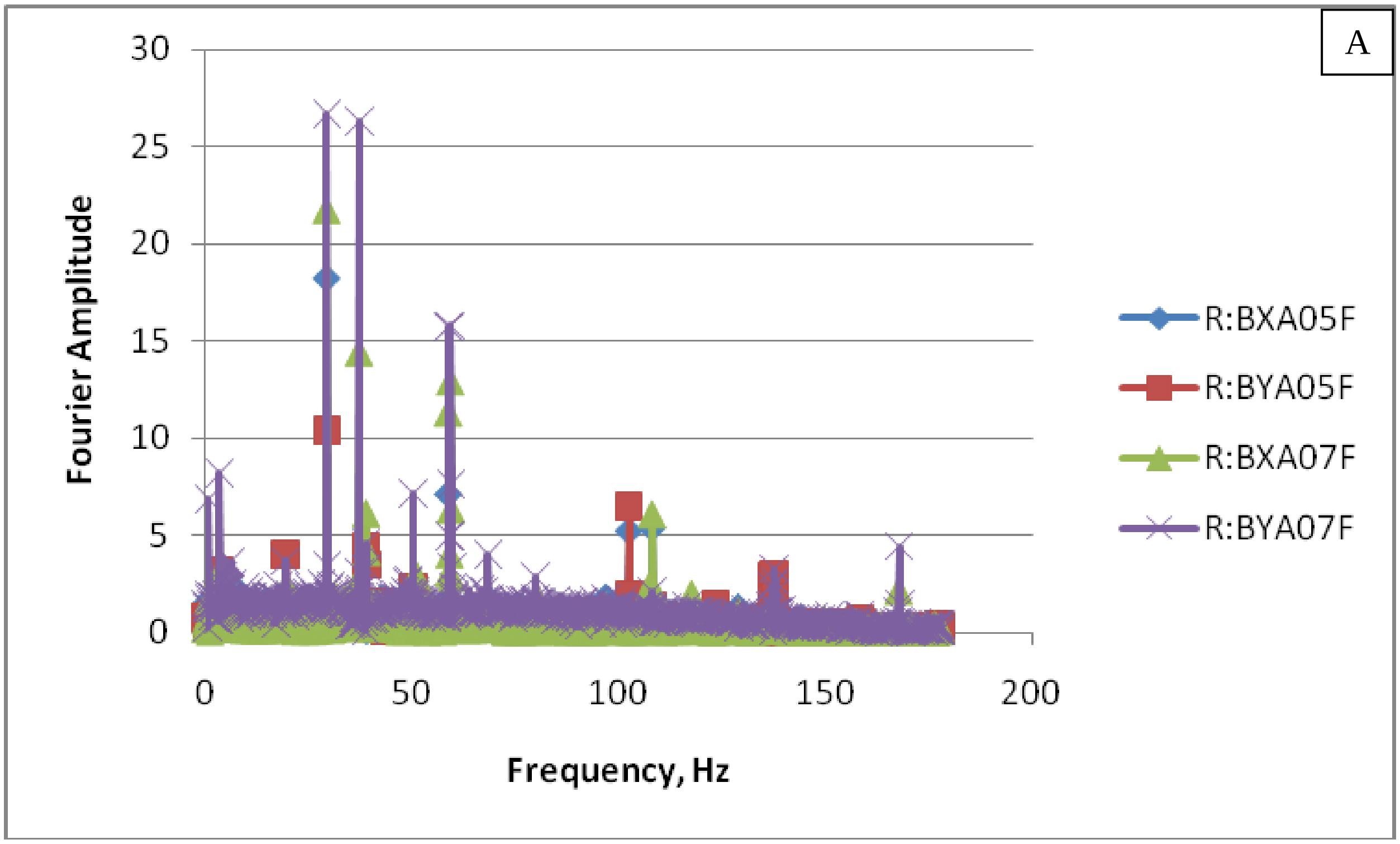

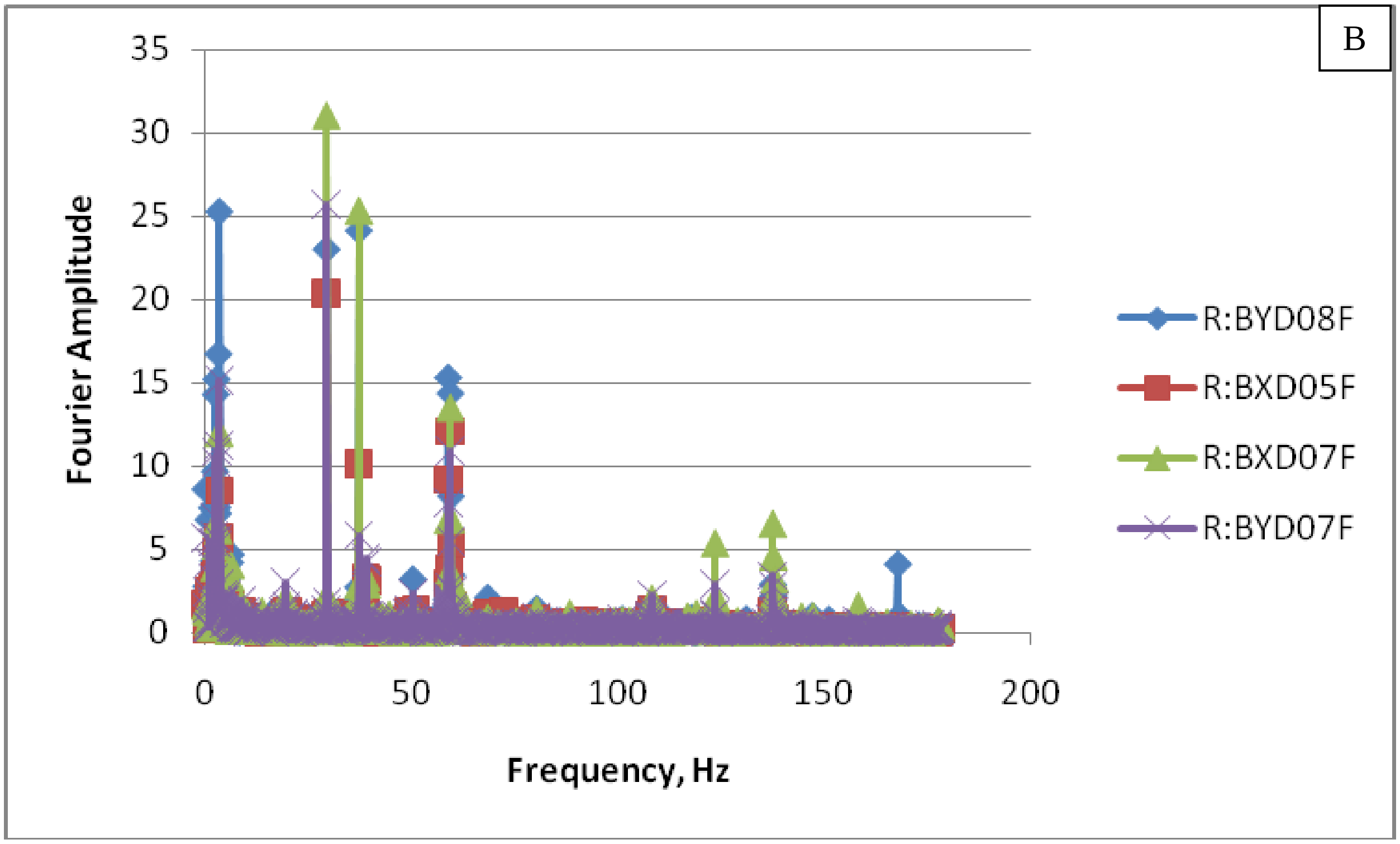

**Figure 6.5**. FFT spectra of 8 BPMs recorded on August 6, 2010. The acquisition frequency was 256 Hz; 2048 points were recorded for each channel. The beam current was 184 mA, Case A. The upper plot presents the BPMs in low-dispersion locations (before the bend), and the lower plot is for high-dispersion locations (after the bend).

### *D. Other beam motion components*

As it has been shown above, a large component of the beam motion is not associated with the energy ripple. The strongest lines are the rotation frequency of the shaft motor, 29.8 Hz, and its second harmonics, and the second harmonics of the chain motor rotation frequency, $2 \times 19.3 = 38.6$ Hz. The contribution can come either from vibrations (discussed in detail in Ref. [6.1]) or can be caused by stray magnetic fields from the motors. Rms values of the BPM signals are summarized in Table 6.2.

**Table 6.2**. Rms values of BPM signals presented in Fig.6.5 (in mm).

| BXA05 | BYA05 | BXA07 | BYA07F | BXD08 | BYD08 | BXD07 | BYD07 | BXD05 | BYD05 |
|---|---|---|---|---|---|---|---|---|---|
| 0.017 | 0.012 | 0.027 | 0.035 | 0.022 | 0.036 | 0.036 | 0.029 | 0.026 | 0.016 |

An accurate estimation of a possible effect of these oscillations on angles in the cooling section requires tracking of specific lines from the spectrum similarly to what was done in Ref. 6.1. To estimate the order of magnitude, we can assume that the beam sizes around A05 location and in the future BNL cooling section are similar, and, correspondingly, the oscillation amplitudes will be similar as well. The resulting angles should be comfortably below of the expected total angle of 0.1 mrad.

## E. Hysteresis in bending magnets

One of the concerns for using Fermilab's bending magnets for low-energy running is the quality of their magnetic fields at low field strength. Two 45 degree magnets are in the U-bend beam line, being turned off for this mode of operation. This allows making an estimate one of the bend's characteristics, its hysteresis. The beam trajectory was recorded before and after cycling the bends from zero to its full nominal current (for 4.3 MeV operation) of ~4A and back to zero. The difference between the before and after trajectories is shown in Fig. 6.6A. Because cycling the bends required cycling the Pelletron as well, data were recorded when cycling the Pelletron only (Fig. 6.6B). In the latter case, the trajectory was reproducible, and the entire difference in Fig. 6.6A should be attributed to the hysteresis in the bends.

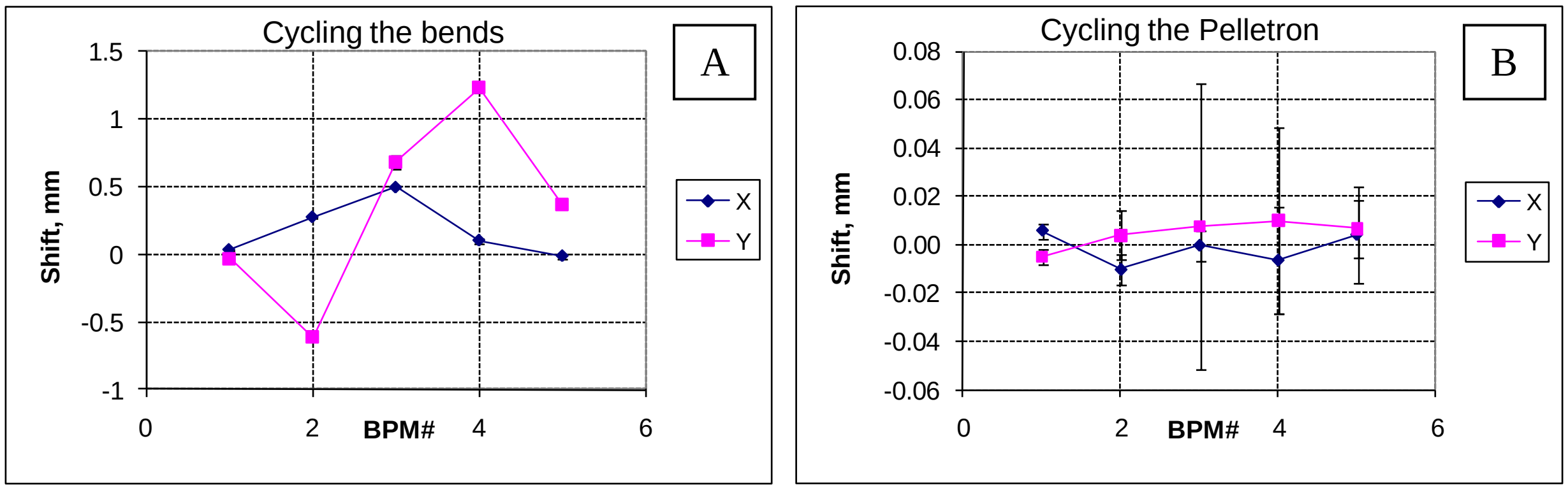

**Figure 6.6.** Difference between trajectories recorded before and after cycling the bends (A) and the Pelletron only (B). August 6, 2010. Each trajectory was recorded for 20 times at 1 Hz Points show the average values, and the error bars are the standard deviations.

Using the OptiM program, the field of the dipoles was fitted to match the resulting orbit to shifts measured in BPMs labeled #2-4 in Fig. 6.6A. The resulting trajectory is shown in Fig. 6.7 and the fitted field integrals are listed in Table 6.3. We do not have a clear explanation for a poor fit in the last BPM.

When the magnets are used at the energy of 1.6 MeV, the observed field perturbation is ~0.5% of the main field. If the field quality is of the same order, it may result in focusing aberrations. Measuring the magnet properties at low bending fields is desirable as well as foreseeing bipolar bend power supplies for degaussing. Note that the effect becomes even a bigger concern for running at lower energies.

**Table 6.3:** Change of the integrated dipole field strength obtained from the change in the orbit.

| | DYS1A | DYT6B |
|---|---|---|
| Integrated field in X direction [Gauss·m] | -0.305 | -0.555 |
| Integrated field in Y direction [Gauss·m] | 0.009 | 0.067 |

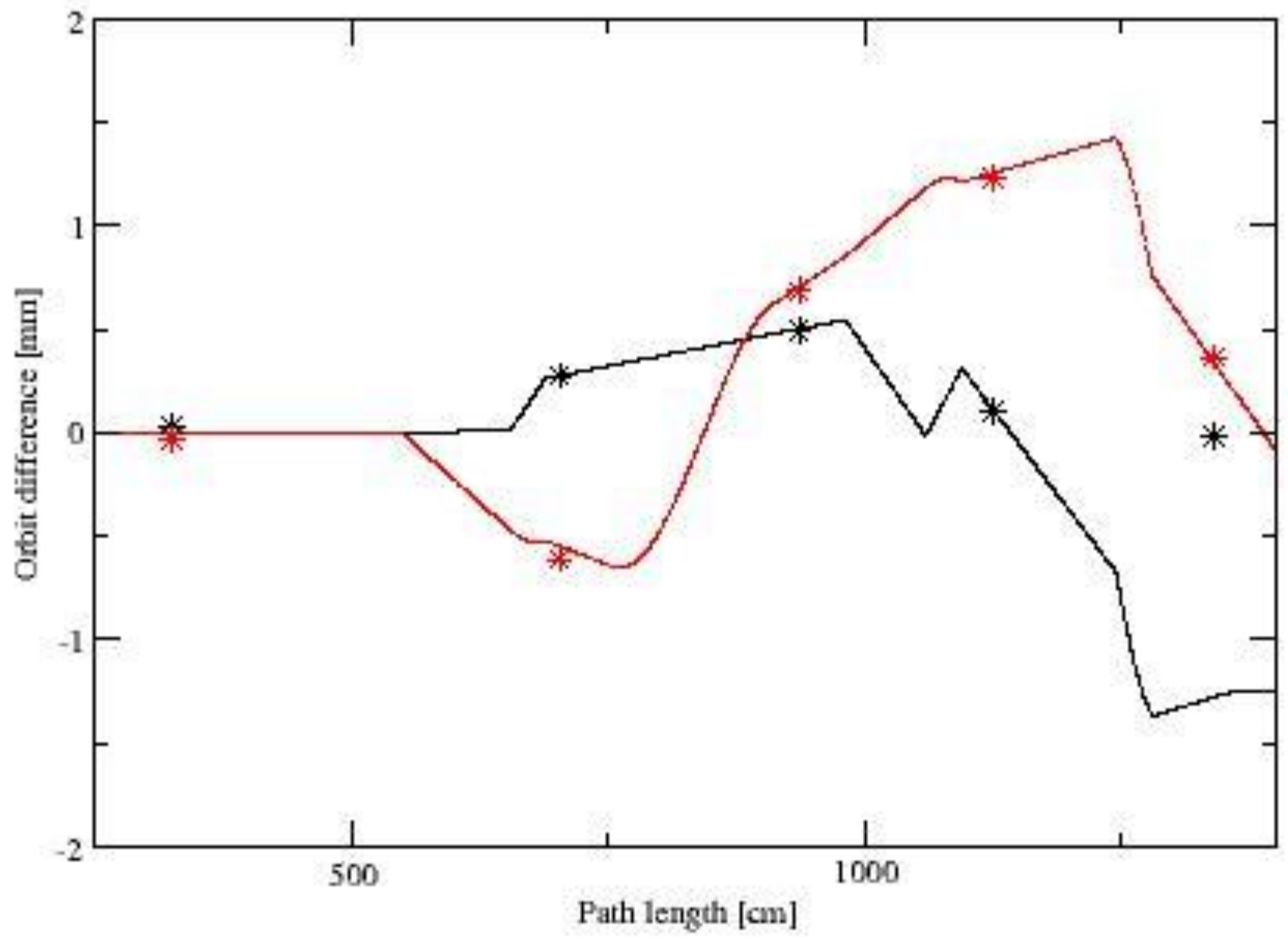

**Figure 6.7** Measured difference orbit (stars) after cycling the DY dipole magnets and OptiM simulation (lines) using orbit kicks in the position of the dipoles. The X direction is shown in black, Y direction is shown in red.

# VII. Summary

The low-energy run of Fermilab's Electron cooler showed that the system of beam generation and energy recovery is capable of operating at 1.6 MeV. The maximum recirculated current was 0.38A, well above what is presently considered, 0.1A. The estimated energy ripple is ~40 eV, which looks satisfactory. The recirculation stability was excellent; during the entire run no unprovoked beam interruptions were observed. As a result of this effect, improved diagnostics and accrued experience, commissioning of this mode of operation was incomparably faster than the original one in 2005. Note that the described mode of operation was conducted with a significantly higher magnetic flux through the cathode, which opens the range of available beam sizes in the cooling section.

Nevertheless, several issues were identified during the run:

- High voltage regulation does not work properly at the level of 1 MV. If extending the operation capabilities of the Pelletron to this region is found to be necessary, modifications (likely, minor) are required.
- At the lower energy, the present protection system based on ionization chambers is inadequate.
- Additional magnetic measurements of the bending magnets are needed to determine at what parameters they can be used in the low-energy mode.

In summary, using the Electron cooler for the BNL low-energy RHIC program is feasible.